\documentclass[]{mn2e}
\newcommand{\feh}{\mbox{[Fe/H]}}
\newcommand{\zh}{\mbox{[Z/H]}}
\newcommand{\afe}{\mbox{[$\alpha$/Fe]}}

\usepackage{graphicx}
\title[Globular clusters in NGC~147, NGC~185, and NGC~205]{Ages, metallicities, and $\afe$ ratios of globular clusters in NGC~147, NGC~185, and NGC~205}
\author[M. E. Sharina, V. L. Afanasiev, and T. H. Puzia]{M. E. Sharina$^{1}$\thanks{E-mail:
sme@sao.ru (MES); vafan@sao.ru (VLA); tpuzia@stsci.edu (THP)},
V. L. Afanasiev$^{1}$, and T. H. Puzia$^{2}$\\
$^{1}$Special Astrophysical Observatory, Russian Academy of Sciences,
N.Arkhyz, KChR, 369167, Russia\\
$^{2}$Space Telescope Science Institute, 3700 San Martin Drive, Baltimore, MD 21218, USA}
\begin{document}

\date{Accepted 2006 August 10. Received 2006 August 07; in original form 2006 March 31}

\pagerange{\pageref{firstpage}--\pageref{lastpage}} \pubyear{2006}

\maketitle

\label{firstpage}

\begin{abstract}
We present measurements of ages, metallicities, and $\afe$ ratios for 16
globular clusters (GC) in NGC~147, NGC~185, and NGC~205 and of the central
regions of the diffuse galaxy light in NGC~185, and NGC~205. Our results
are based on spectra obtained with the SCORPIO multi-slit spectrograph at
the 6-m telescope of the {\it Russian Academy of Sciences}. We include in
our analysis high-quality HST/WFPC2 photometry of individual stars in the
studied GCs to investigate the influence of their horizontal branch (HB)
morphology on the spectroscopic analysis. All our sample GCs appear to be
old ($T>8$ Gyr) and metal-poor ($\zh \la -1.1$), except for the GCs
Hubble~V in NGC~205 ($T=1.2\pm0.6$ Gyr, $\zh=-0.6\pm0.2$), Hubble~VI in
NGC~205 ($T=4\pm2$ Gyr, $\zh =-0.8\pm0.2$), and FJJVII in NGC~185
($T=7\pm3$ Gyr, $\zh =-0.8\pm0.2$). The majority of our GC sample has
solar $\afe$ enhancement in contrast to the halo population of GCs in M31
and the Milky Way. The HB morphologies for our sample GCs follow the same
behavior with metallicity as younger halo Galactic globular clusters. We
show that it is unlikely that they bias our spectroscopic age estimates
based on Balmer absorption line indices. Spectroscopic ages and
metallicities of the central regions in NGC~205 and NGC~185 coincide with
those obtained from color-magnitude diagrams. The central field stellar
populations in these galaxies have approximately the same age as their
most central GCs (Hubble~V in NGC~205 and FJJIII in NGC~185), but are more
metal-rich than the central globular clusters.
\end{abstract}


\section{Introduction}
Understanding the role of dwarf galaxies and their globular cluster
systems as building blocks of massive early-type and spiral galaxies is
one of the great challenges of modern astrophysics. Studying the chemical
compositions and color-magnitude diagrams (CMDs) of globular clusters
(GCs) in the nearby low-mass galaxies is critical to compare properties of
these long-living objects situated in galaxies of different type and mass.
Only the Local group (LG) dwarf galaxies are close enough so that their
globular clusters can be resolved into single red giant and horizontal
branch stars by today's observatories.

Three close companions of the Andromeda galaxy, NGC~147, NGC~185 and
NGC~205, comprise the brightest end of the LG dwarf early-type galaxy
luminosity function. The proximity to M31 in combination with the unusual
properties of these three galaxies have attracted great attention of
astronomers since the beginning of the 20th century (we refer the reader
to a monograph by Van den Bergh 2000 and a review by Mateo 1998 for
details). NGC~205 and NGC~185 are very unusual galaxies and do not
resemble any galaxy within $\sim\!10$ Mpc. Both have a regular spheroidal
shape and are composed mainly of old stars. At the same time, both
galaxies have a considerable amount of gas, dust, and a significant
intermediate-age stellar component.  These unique properties are most
likely tightly related their unusual evolutionary histories (e.g. Yong \&
Lo, 1997; Davidge, 1992; Lee et al., 1993; Butler \& Martinez-Delgado,
2005; Davidge, 2005). Compared to its brighter cousins, NGC~147 is gas and
dust free. Asymptotic giant branch (AGB) stars contribute only 2-3\% of
the total light of this galaxy (Davidge, 1994). By means of its stellar
content, NGC~147 resembles a typical dwarf spheroidal galaxy in the Local
Group. However, it is much more luminous. The distance to NGC~205, NGC~185
and NGC~147, their stellar populations and star formation histories are
being studied actively (e.g.~McConnachie 2005, Butler \& Martinez-Delgado
2005, Dolphin 2005). Table~1 summarises their distances, brightnesses, and
reddenings used in this paper for the studied galaxies. We use a standard
reddening law with $A_V = 3.315\cdot E(B-V)$. We note that we
find indications for a slightly smaller distance modulus, $(m-M)_0=24.38 \pm 0.1$ mag,
for NGC~205 compared to the values derived by McConnachie (2005),
$(m-M)_0= 24.58 \pm 0.07$ mag, and Dolphin (2005), $(m-M)_0=24.45 \pm 0.14$ mag,
based on the results of our stellar photometry (see Sect.~3).
\begin{table}
\centering
\scriptsize
\caption{Absolute magnitudes, distance moduli and reddenings
from Schlegel et al. (1998)
used for the galaxies of our study (indices refer to 0: this work; 1:
McConnachie et al. (2004).}
\begin{tabular}{@{}lcccc@{}}
\hline
NGC   & $M_{V,0}$ & $(M-m)_0$ &    & $A_V$        \\
      &           &           &         \\ \hline
147   & $-15.1^1$ & 23.95$^1$ &  0.580  \\
185   & $-15.6^1$ & 24.15$^1$ &  0.593  \\
205   & $-16.2^0$ & 24.38$^0$ &  0.206  \\
\hline
\end{tabular}
\end{table}

\begin{table}
\centering
\scriptsize
\caption{Properties of globular clusters in NGC~147, 185 and 205 summarised from
the literature. Columns contain the following data: (1) object name,
(2) integrated V magnitude, (3) integrated $U-B$ colour, (4) integrated $B-V$
colour, (5) metallicity (indices refer to
1: Hodge (1974), 2: Hodge (1976), 3: Hodge (1973)
4: Da Costa and Mould (1988), 5: Geisler et al. (1999),
6: Butler and Martinez-Delgado (2005), 7: Sharov and Lyuty (1983),
0: our HST/WFPC2 integrated $VI$ photometry
corrected for Galactic extinction using Schlegel et al. (1998) maps.
 A star like symbol indicates, that $(V-I)_0$ color is given instead of $(B-V)$.}
\begin{tabular}{@{}lcllc@{}}
\hline
Object     & V         & $U-B$    &$B-V$   & \feh\      \\
\hline
{\bf NGC~147}&         &          &        &                  \\
HodgeI     & $17.7^2$  & $0.6^2$  &$0.8^2$ & $-1.9^4$ \\
HodgeII    & $16.5^1$  & $0.4:^7$ &$0.6^7$ &  .... \\
HodgeIII   & $17.0^2$  & $0.1^2$  &$0.6^2$ & $-2.5^4$ \\
	   &           &          &        &      \\
{\bf NGC~185}&         &          &        &      \\
FJJI       & $18.4^1$  & $0.2^1$  &$0.8^1$ & $-1.4^4$ \\
FJJII      & $19.7^1$  & ....     &....    & $-1.2^4$ \\
FJJIII     & $16.8^1$  & $0.1^1$  &$0.9^1$ & $-1.7^4$, $-1.6^5$ \\
FJJIV      & $19.0^4$  & $0.05^1$ &$0.7^1$ & $-2.5^4$, $-1.9^5$ \\
FJJV       & $16.7^1$  & ....     &$0.7^1$ & $-1.8^4$, $-1.5^5$ \\
FJJVII     & $18.9^{0,*}$ & ....  &$1.0^{0,*}$ &  .... \\
	   &           &          &        &      \\
{\bf NGC~205}&         &          &        &      \\
HubbleI    & $16.9^3$  &  ....    &$0.8^3$ & $-1.5^4$ \\
HubbleII   & $16.7^3$  & $0.2^3$  &$0.9^3$ & $-1.5^4$ \\
HubbleV    & $16.7^3$  & $0.2:^3$ &$0.6^3$ &   .... \\
	   &           &          &        &    .... \\
HubbleVI   & $17.9^3$  & $0.8^3$  &$0.7^3$ & $-1.3^4$ \\
HubbleVII  & $18.0^3$  & $0.1^3$  &$0.9^3$ & $-1.4^4$ \\
HubbleVIII & $16.6^3$  & $0.1^3$  &$0.7^3$ & $-1.9^4$  \\
\hline
\end{tabular}
\end{table}
\begin{table}
\centering
\scriptsize
\caption{Journal of spectroscopic observations}
\begin{tabular}{@{}lclr@{}}
\hline
Object      & Date       & Exposure & Seeing         \\ \hline
{\bf NGC~147}      &            &          &                 \\
Hodge I, II, III   & 23.09.2003 & 3x900s   & 1.2       \\
\noalign{\smallskip}
%II, III     &            &          &           \\
{\bf NGC~185}      &            &          &                 \\
FJJIII, IV, V, VI, VII& 22.09.2003 & 4x900s   & 1.2       \\
% V, VI, VII &            &          &           \\
FJJI, FJJII & 26.09.2003 & 3x900s   & 1.2       \\
FJJVIII     & 26.09.2003 & 3x600s   & 1.2       \\
\noalign{\smallskip}
{\bf NGC~205}     &            &          &                 \\
Hubble II, V, VI, VII & 23.09.2003 & 4x900s   & 1.2       \\
%V, VI, VII  &            &          &                 \\
Hubble I    & 25.09.2003 &   900s   & 1.2       \\
%	    &            &          &           \\
\noalign{\smallskip}
\noalign{\smallskip}
{\bf Stars}      &            &          &                 \\
HD10307  & 25.09.2003 & 2x30s      & 1.2     \\
HD26784  & 26.09.2003 & 2x20s      & 1.2     \\
HD26757  & 26.09.2003 & 2x20s      & 1.2     \\
%	 &            &         &           \\
BD+26377 & 26.09.2003 & 2x30s   &  1.2      \\
\hline
\end{tabular}
%\end{minipage}
\end{table}

Each of our sample galaxies hosts a sizable globular cluster system (GCS).
The discovery of globular clusters in NGC~147, NGC~185 and NGC~205
includes the work of Hubble (1932), Baade (1944), Hodge (1974) and Ford,
Jacoby, \& Jenner (1977, hereafter FJJ77) (see Da Costa and Mould (1988)
and van den Bergh (2000) for details). All searches of GCs were carried
out on photographic plates. The clusters were classified as being
associated with the galaxies by their galactocentric radius. However,
recent studies of GCSs in dwarf galaxies reveal globular clusters situated
far from optical bodies of their host galaxies (see e.g. Hwang et al.
2005). It is likely that detailed wide-field investigations may discover
additional GC candidates around NGC~147, NGC~185 and NGC~205 using new
observational facilities and improved classification schemes which would
allow the measurement of real physical sizes of the GCSs in these dwarf
galaxies. Four extended globular clusters were discovered in the outer
halo of M31 by Huxor et al., 2005) using INT-WFS images. Recent HST based
surveys reveal new globular cluster candidates in M31 (Barmby \& Huchra
2001, see also the summary of optical and IR photometry in Galleti et al.
2004). For instance, there are $\sim\!50$ objects in the catalog of
Galleti et al. (2004) within a radius of 20\farcm\ around the center of
NGC~205. Galleti et al. (2006) emphasize that our knowledge about the M31
GCS is far from complete even in terms of simple membership and undertake
a large spectroscopic survey of bona-fide globular clusters and cluster
candidates in M31. Obtaining radial velocities for GCs in the M31 halo and
in the dwarf galaxies, detailed CMD and abundance studies will certainly
help to solve the problem of the origin of M31 and of the satellite
galaxies.

The properties of the globular clusters in NGC~147, NGC~185 and NGC~205
from the literature are summarised in Table~2. Da Costa and Mould (1988)
first obtained spectra for eight Hubble clusters, the nucleus of NGC~205,
FJJ77 clusters I--V in NGC~185 and for the two clusters Hodge~I and III in
NGC~147. A comparison of line strengths relative to similar spectra of
Galactic globular clusters indicated that with a single exception, of the
central GC Hubble~V in NGC~205, all the clusters in NGC~147, NGC~185 and
NGC~205 are old and metal-poor. The age estimates for Hubble~V fall in a
wide range from $\sim\!300$ Myr (based on $BVR$ photometry, Lee, 1996) to
intermediate ages (based on optical and far-ultraviolet imaging with
HST/WFPC2, Jones et al., 1996). Star cluster formation seems to be a more
extended process in NGC~185. A star forming region of $H \alpha+[NII]$
emission with a diameter of ~50 pc near the center of NGC~185 may be a
probable place of a massive young star cluster birth according to a modern
evolutionary scheme (Gallagher \& Grebel, 2001 and references therein).
Young \& Lo (1997) estimated the mass of the brightest central HI clump in
NGC~185 $M_{HI}=0.7 \cdot 10^4 M_{\sun}$ and the corresponding virial mass
a factor of 100 higher. The clump is associated with a massive giant
molecular cloud.

\begin{figure*}
\vspace{-2cm}
\includegraphics[width=0.9\textwidth]{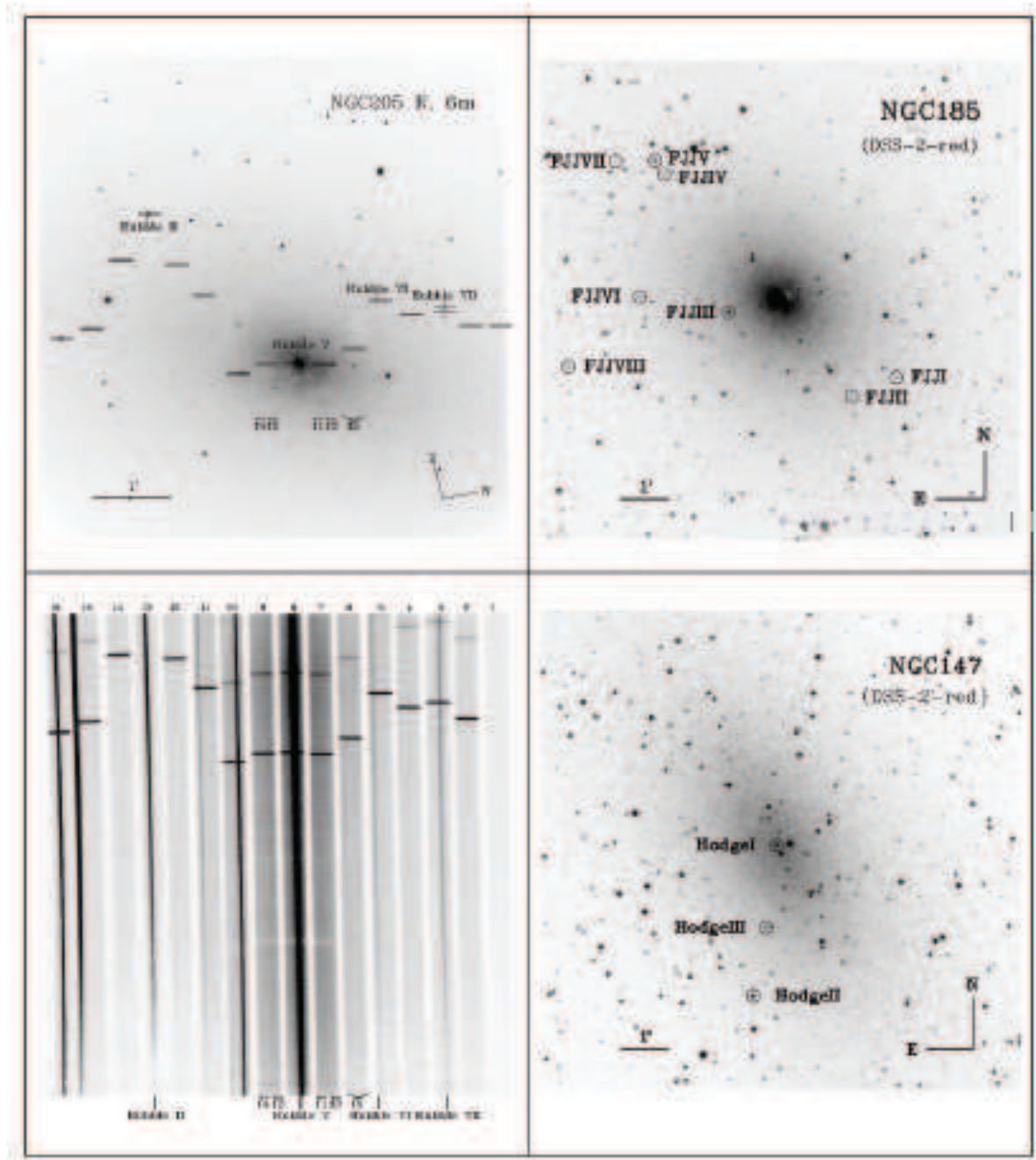}
\vspace{-2cm}
\caption{The upper left panel shows an image of NGC~205 obtained 
with the SCORPIO multi-slit unit at the 6-m RAS telescope. Settings of slits on the
globular clusters and diffuse galactic light regions are indicated. The
lower left panel shows a frame with raw non-reduced spectra of objects in
NGC~205. Two right panels show Digital Sky Survey 10\arcmin\ by 10\arcmin\
images of NGC~185 and NGC~147 with globular clusters.}
\label{dss}
\end{figure*}

In this paper, we derive ages, metallicities, and \afe\ ratios for
globular clusters in NGC~147, NGC~185 and NGC~205 based on measurements of
absorption line indices as defined by the Lick standard system (Burstein
et al. 1984; Worthey et al. 1994; Worthey \& Ottaviani 1997; Trager et al.
1998). The main advantage of this system is its ability to disentangle age
from metallicity using the information on the strength of prominent
diagnostic absorption line features measured in the
intermediate-resolution integrated-light spectra.

The paper is organised as follows. In Section~2 we describe our
spectroscopic data, the data reduction, as well as our methods and results
of measuring Lick line indices in the spectra of globular clusters and the
diffuse host galaxy light. In Section~3 we present results from our CMD
photometry analysis of globular clusters, where we compare the loci of red
giant branch (RGB) and horizontal branch (HB) stars with the
Victoria-Regina stellar models of VandenBerg et al. (2006). In Section~4 we
discuss the derived evolutionary parameters for globular clusters and
compare them with the data available in the literature.

\section{Spectroscopic data}
\subsection{Observations and data reduction}
The observations were performed with the multi-slit unit of the SCORPIO
spectrograph (Afanasiev \& Moiseev, 2005), installed at the prime focus of
the 6-m telescope of the {\it Russian Academy of Sciences}. Fig.~\ref{dss}
shows DSS-2-red images of NGC~147 and NGC~185 with marked globular
clusters as well as the 6-m telescope image of NGC~205 with marked
settings of the slits and an image of the corresponding raw spectra.
In multi-slit mode SCORPIO has 16 movable slits (1\farcs2 x 18\farcs0) in the
field of 2\farcm9 x 5\farcm9 in the focal plane of the telescope. We use
the CCD detector EEV42-40 with 2048 x 2048 pixel elements and the scale
$\sim\!0\farcs18$ per pixel. The grism GR1200g (1200 lines/mm) gives a
spectral resolution $\sim\!5$ \AA. The exact spectral coverage of an
individual spectrum depends on the distance of the slit position from the
meridian of the field. Most of our spectra have good wavelength coverage
between $4200-5800$ \AA. Slit positions were set based on pre-imaging
obtained with the same instrument.

\begin{figure}
\includegraphics[width=0.48\textwidth]{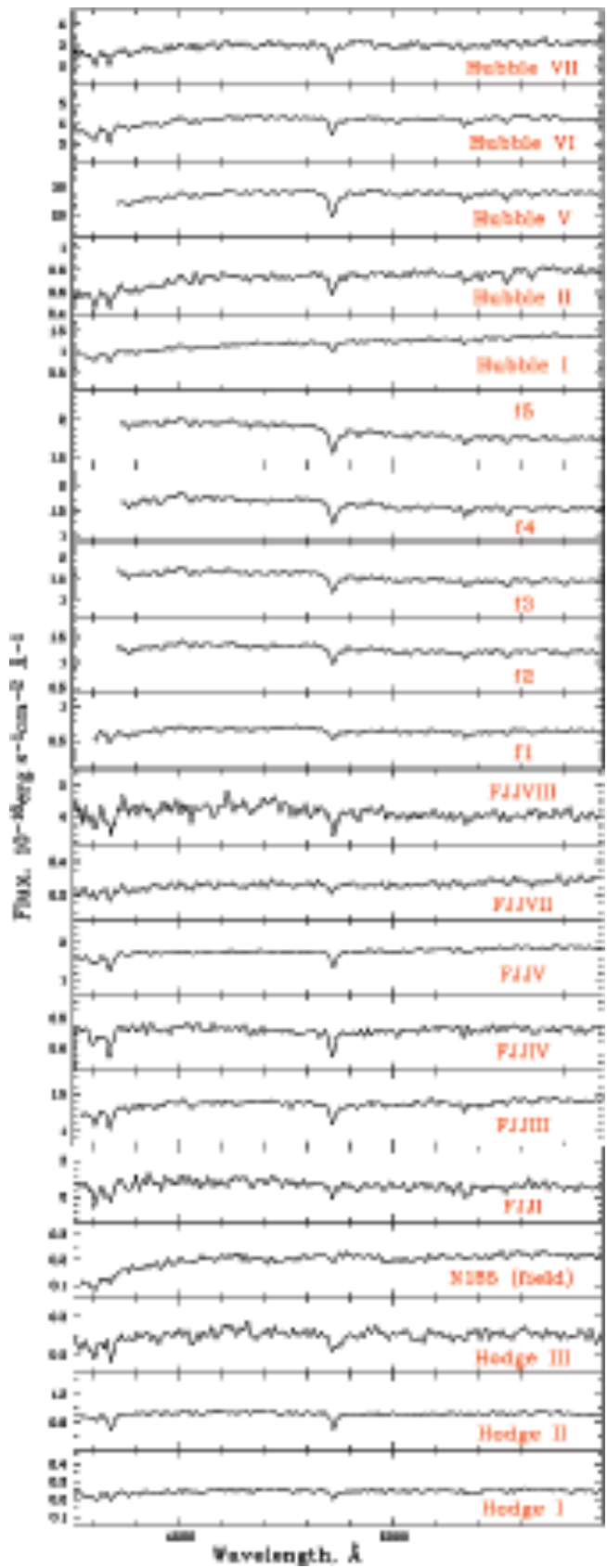}
\caption{Flux calibrated spectra of GCs and diffuse light in
NGC~147, NGC~185 and NGC~205 at a constant resolution of 8 \AA\ at all wavelengths.}
\label{spec_all}
\end{figure}

A journal of observations is given in Table~3. Globular clusters covered
by one slitmask pointing are listed in one row. All objects, including
standard stars, were observed with the same setup of the multi-slit unit.
Resulting spectra of the globular clusters and of diffuse light are shown
in Fig.~\ref{spec_all}.
\begin{table}
\centering
\scriptsize
\caption{Correction terms of the transformation to the Lick/IDS
standard system Worthey et al. (1994). We applied them in the
following way $ I_{\rm Lick}=I_{\rm measured}+c $.
 The range of index values covered by our standard stars is shown
in the fourth column.}
\begin{tabular}{lrccr} \\ \hline
Index       & c          & rms error & Index range& units \\ \hline
Ca4227      & 0.117      & 0.055&   [0.03 -- 0.8]    &  \AA   \\
G4300       & 0.520      & 0.332&   [2.6 -- 5.3]     &  \AA   \\
Fe4384      & 0.767      & 0.235&   [2.3 -- 4.0]     &  \AA   \\
Ca4455      & 0.851      & 0.136&   [0.9 -- 1.3]     &  \AA   \\
Fe4531      & 0.197      & 0.611&   [2.2 -- 2.7]     &  \AA   \\
Fe4668      & $-$0.068   & 0.673&   [2.2 -- 3.8]     &  \AA   \\
H$\beta$    &$-$0.024    & 0.197&   [2.3 -- 3.8]     &  \AA   \\
Fe5015      & 0.577      & 0.284&   [3.4 -- 4.2]     &  \AA   \\
Mg$_1$      & 0.009      & 0.008&   [0.01 -- 0.03]   &  mag    \\
Mg$_2$      & 0.019      & 0.006&   [0.08 -- 0.16]   &  mag    \\
Mgb         & 0.243      & 0.162&   [1.9 -- 3.0]     & \AA  \\
Fe5270      & 0.223      & 0.546&   [1.5 -- 2.3]     & \AA  \\
Fe5335      & 0.201      & 0.398&   [1.4 -- 1.7]     & \AA  \\
Fe5406      &  0.120     & 0.049&   [0.7 -- 1.2]     & \AA  \\
H$\gamma_A$ & $-$0.417   & 0.383&   [0.4 -- 2.4]     & \AA  \\
H$\gamma_F$ &  $-$0.191  & 0.364&   [0.0 -- 2.0]     &  \AA  \\
\hline
\end{tabular}
\end{table}
The detailed description of all reduction steps
is given in Sharina et al. (2005). Here we briefly
review major points of our reduction strategy.
The basic data reduction was carried out using software packages within
the {\it Interactive Data Language} (IDL). The subsequent data analysis
was carried out in MIDAS. The data reduction included removing cosmics,
bias subtraction, flat-field correction, and correction of geometric field
distortion. After wavelength calibration and sky subtraction, aperture
windows in the direction of the spatial axis were defined, and the spectra
were extracted by summing the two-dimensional spectra in the spatial
direction. Then the spectra were corrected for atmospheric extinction and
were flux-calibrated. We use spectroscopic Lick standard stars, observed
in the same night through the central slit of the slit mask, for flux
calibration of all spectra. In those nights when radial velocity standard
stars were not observed, we use spectra of the twilight sky to compute
radial velocities relative to telluric lines. In addition to the spectra
of globular clusters, we analysed spectra of the diffuse galaxy light in
the same way. These spectra were obtained near the cluster FJJIII in
NGC~185 and in fields f1 -- f5 in NGC~205 (see Fig.~\ref{spec_all} for details).
\begin{figure*}[!th]
\begin{center}
\includegraphics[width=1.\textwidth]{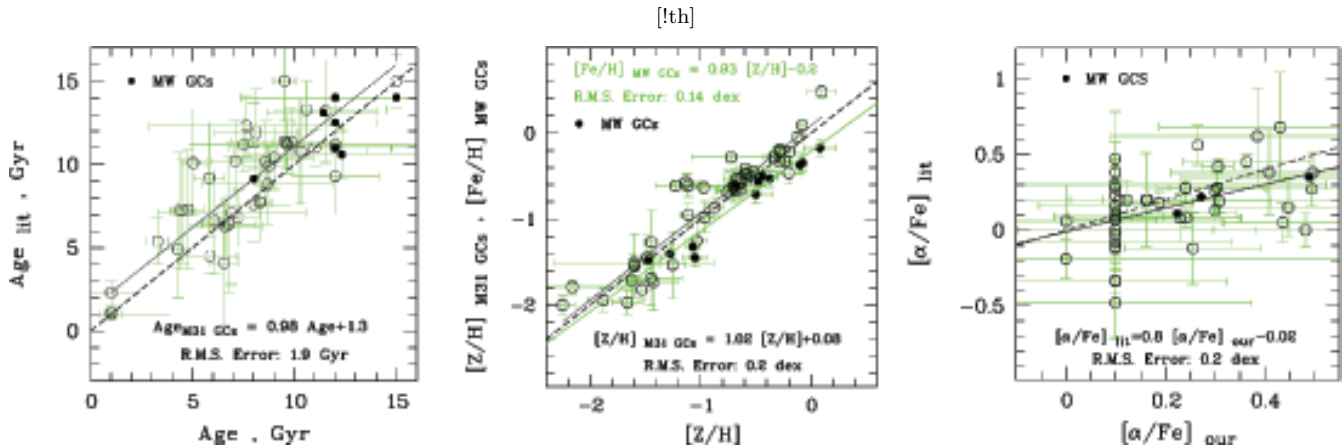}
\end{center}
\caption{Age, \zh, and \afe\ ratios obtained with our fitting routine (see Section 2.3)
for Galactic globular clusters (GCCs) (filled circles)
using the Lick index data of Puzia et al. (2002)
are compared with the ages obtained from CMD studies (Salaris \& Weiss, 2002),
\feh\ data from Harris (1996), and \afe\ values from Pritzl et al. (2005).
Age, \zh\, and \afe\ obtained in this work with our code for M31 GCS (open circles) 
using the Lick index data from Worthey et al. (1994), Kuntschner et al. (2002), Beasley et al. (2004) and Puzia et al. (2005a) are compared with ages, \zh\, and \afe\
 from Puzia et al. (2005a). Dashed lines show the one-to-one relation.
Solid lines indicate the least-square fit to the data.}
\label{compar}
\end{figure*}

\subsection{Calibration of Lick indices}
Before measuring indices, we degraded our spectra to the resolution of the
Lick system, taking into account its wavelength-dependent resolution
(Worthey \& Ottaviani, 1997). The effective resolution of our spectra has
been determined using spectra of standard stars as a width (FWHM) of the
corresponding autocorrelation function divided by $ \sqrt 2$ (Tonry \&
Davis 1979). The resolution-correction technique is described in detail in
Puzia et al. (2002). Lick indices were measured on spectra corrected for
radial velocities (Table~7), with a typical radial velocity error of $\sim
30$ km s$^{-1}$.  Chilingarian et al. (2005) measured radial velocities
for GCs FJJIV and HodgeII according to our request using his method and
obtained $V_h=-182 \pm 12$ and $V_h=-128 \pm 9$ km s$^{-1}$,
correspondingly, in very good agreement with our measurements. The spectra
of diffuse galactic light were corrected for velocity dispersion
determined using algorithms of Tonry \& Davis (1979). Broadening of lines
in the spectra of diffuse galactic light due to velocity dispersion for
the fields f1 -- f5 in NGC~205 and the field in NGC~185, accordingly, was
obtained to be 1.5 $\pm$0.3 \AA, 1.5 $\pm$0.3 \AA, 1.4 $\pm$0.3 \AA, 1.3
$\pm$0.3 \AA, 1.3 $\pm$0.3 \AA, 1.1 $\pm$0.3 \AA, 0.9 $\pm$0.5 \AA. The
velocity dispersion determined by us for the central parts of the galaxies
is slightly larger than that from the literature (Carter \& Sadler 1990,
Bender et al. 1991, Prugniel \& Soubiran 2001), but still well within the
errors. Table~4 summarizes zero points of the linear transformation of our
index measurements to the Lick/IDS standard system with the corresponding
rms errors. We obtained these zeropoints from observations of standard
stars listed in Table~3. These zeropoints coincide well within the errors
with those obtained in other nights from observations of a large set of
Lick standard stars with the multi-slit unit of the SCORPIO spectrograph
equipped with a grism VPHG1200g which provides a spectral resolution
$\sim\!5$ \AA\ (Sharina et al., 2006). Lick indices were measured using
the code GONZO (Puzia et al. 2002). The Lick index uncertainties were
determined from bootstrapping the object spectrum. All index values and
the errors for observed globular clusters, diffuse-light fields, and
combined spectra of globular clusters are presented in Tables~5 and 6.

\subsection{Ages, metallicities and [$ \alpha$/Fe] ratios}
In the following we compute ages, metallicities, \zh\ \footnote{We use the
standard definition, $[X/Y]=log(X/Y)- log(X_{\sun}/Y_{\sun})$,
where X and Y are masses of specific elements. A designation
\zh\ implies the overall metallicity.},
and [$\alpha$/Fe] for our
sample globular clusters using the information acquired from the
measurements of the Lick indices.
\begin{table*}
\begin{center}
\scriptsize
\caption{Globular cluster indices ($ \lambda <$ 4700 \AA) (first line)
corrected for zeropoints of transformation to the standard system (Table~4)
and errors determined from bootstrapping the object spectrum (second line).
Approximate S/N ratios per pixel measured at 5000\AA\ of the initial one-dimensional spectra, not degraded to the resolution of the Lick system are listed in the second column.}
\begin{tabular}{lcrrrrrrrr} \\ \hline \hline
ID               &  S/N &    Ca4227& G4300      & H$ \gamma_A$& H$ \gamma_F$&      Fe4383 &      Ca4455&      Fe4531 &      Fe4668 \\
		 &      &    (\AA) & (\AA)      &      (\AA)&        (\AA)  &      (\AA)  &      (\AA) &       (\AA) &       (\AA) \\ \hline
HodgeI           & 20   &    0.15  &      3.41  &       0.10&        1.27   &       2.53  &      0.42  &       0.91  &       0.16  \\
		 &      & $\pm$0.05&  $\pm$0.13 &  $\pm$0.42 &  $\pm$0.42 &  $\pm$0.19 &  $\pm$0.20 &  $\pm$0.24 &  $\pm$0.31 \\
HodgeII          & 34   &    0.36  &      0.58  &       2.64&        2.04   &       0.55  &      0.79  &       0.86  &      -0.44  \\
		 &      & $\pm$0.02&  $\pm$0.06 &  $\pm$0.25 &  $\pm$0.25 &  $\pm$0.10 &  $\pm$0.11 &  $\pm$0.13 &  $\pm$0.18 \\
HodgeIII         & 14   &    0.14  &      2.30  &       0.72&        1.60   &       0.12  &      0.53  &       1.62  &       1.94  \\
		 &      & $\pm$0.08&  $\pm$0.21 &  $\pm$0.70&  $\pm$0.71 &  $\pm$0.29 &  $\pm$0.32 &  $\pm$0.37 &  $\pm$0.51 \\
all N147 GCs     & 43   &    0.30  &      1.39  &       1.35&        1.80   &       1.05  &      0.66  &       1.08  &      -0.01  \\
		 &      & $\pm$0.02&  $\pm$0.04 &  $\pm$0.15&  $\pm$0.15 &  $\pm$0.06 &  $\pm$0.07 &  $\pm$0.08 &  $\pm$0.11 \\
FJJI             & 18   &    0.84  &      3.00  &       2.78&        2.27   &       2.11  &      0.77  &       0.82  &      -1.41  \\
		 &      & $\pm$0.09&  $\pm$0.20 &  $\pm$0.62&  $\pm$0.63 &  $\pm$0.32 &  $\pm$0.34 &  $\pm$0.37 &  $\pm$0.45 \\
FJJIII           & 55   & ....     &      0.92  &       0.88&        1.65   &       0.51  &      0.74  &       1.11  &      -0.21  \\
		 &      & ....     &  $\pm$0.05 &  $\pm$0.15&  $\pm$0.16 &  $\pm$0.07 &  $\pm$0.08 &  $\pm$0.08 &  $\pm$0.12 \\
FJJIV            & 30   & ....     &      1.02  &       2.32&        2.17   &       0.69  &      0.84  &      -0.28  &       0.36  \\
		 &      & ....     &  $\pm$0.07 &  $\pm$0.29&  $\pm$0.29 &  $\pm$0.12 &  $\pm$0.13 &  $\pm$0.16 &  $\pm$0.21 \\
FJJV             & 70   & ....     &      1.75  &       2.05&        2.06   &       0.93  &      1.15  &       1.21  &      -0.04  \\
		 &      & ....     &  $\pm$0.03 &  $\pm$0.12&  $\pm$0.13 &  $\pm$0.05 &  $\pm$0.06 &  $\pm$0.07 &  $\pm$0.10 \\
FJJVII           & 16   & 0.74     &      2.55  &       0.33&        1.60   &       3.37  &      1.59  &       3.15  &       0.45  \\
		 &      & $\pm$0.10&  $\pm$0.20 &  $\pm$0.72&  $\pm$0.73 &  $\pm$0.31 &  $\pm$0.33 &  $\pm$0.39 &  $\pm$0.58 \\
FJJVIII          & 20   &   -0.12  &      1.61  &       2.08&        1.91   &       1.12  &      0.46  &       1.06  &       0.70  \\
		 &      & $\pm$0.07&  $\pm$0.20 &  $\pm$0.63&  $\pm$0.64 &  $\pm$0.34 &  $\pm$0.36 &  $\pm$0.40 &  $\pm$0.51 \\
NGC185 field     & 27   & ....     &      ....  &      .... &        1.21   &       0.88  &      1.76  &       0.97  &       0.02  \\
		 &      & ....     &      ....  &      .... &   $\pm$0.06&  $\pm$0.21 &  $\pm$0.24 &  $\pm$0.28 &  $\pm$0.36 \\
FJJI+III+IV+     & 106  & ....     &      ....  &       1.70&        2.05   &       1.22  &      0.68  &       1.21  &      -0.54  \\
		 &      & ....     &      ....  &  $\pm$0.06&  $\pm$0.06 &  $\pm$0.02 &  $\pm$0.03 &  $\pm$0.03 &  $\pm$0.04 \\
all N147 GC+     & 120  & ....     &      ....  &       1.61&        2.00   &       1.17  &      0.67  &       1.11  &      -0.47  \\
+FJJI+III+IV+    &      &          &            &  $\pm$0.05 &  $\pm$0.05 &  $\pm$0.02 &  $\pm$0.02 &  $\pm$0.03 &  $\pm$0.03 \\
+V+VIII          &      &          &            &           &               &             &            &             &             \\
HubbleI          & 20   &    0.70  &      2.59  &       0.10&        1.31   &       1.44  &      0.72  &       1.19  &       1.24  \\
		 &      & $\pm$0.08&  $\pm$0.18 &  $\pm$0.64 &  $\pm$0.65 &  $\pm$0.29 &  $\pm$0.30 &  $\pm$0.33 &  $\pm$0.46 \\
HubbleII         & 86   &    0.33  &      2.47  &       0.53&        1.53   &       1.82  &      1.14  &       1.82  &       0.08  \\
		 &      & $\pm$0.00&  $\pm$0.02 &  $\pm$0.07 &  $\pm$0.07 &  $\pm$0.03 &  $\pm$0.04 &  $\pm$0.04 &  $\pm$0.05 \\
HubbleV          & 180  & ....     &      ....  &      .... &        ....   &       0.79  &      1.15  &       1.66  &       1.18  \\
		 &      & ....     &      ....  &      ....  &      ....  &  $\pm$0.00 &  $\pm$0.01 &  $\pm$0.01 &  $\pm$0.01 \\
HubbleVI         & 26   & 0.62     &      2.25  &       2.22&        2.61   &       3.05  &      1.23  &       2.34  &       0.64  \\
		 &      & $\pm$0.05&  $\pm$0.11 &  $\pm$0.35 &  $\pm$0.35 &  $\pm$0.18 &  $\pm$0.19 &  $\pm$0.21 &  $\pm$0.26 \\
HubbleVII        & 36   &    0.47  &      2.68  &      -0.03&        1.42   &       1.52  &      0.55  &       1.76  &      -1.21  \\
		 &      & $\pm$0.03&  $\pm$0.07 &  $\pm$0.23 &  $\pm$0.23 &  $\pm$0.11 &  $\pm$0.12 &  $\pm$0.14 &  $\pm$0.17 \\
HubbleI+VII      & 100  &    0.56  &      2.73  &       0.16&        1.35   &       1.48  &      0.56  &       1.92  &       1.04  \\
		 &      & $\pm$0.02&  $\pm$0.05 &  $\pm$0.13 &  $\pm$0.14 &  $\pm$0.08 &  $\pm$0.08 &  $\pm$0.09 &  $\pm$0.11 \\
HubbleVI+FJJVII  &  30  & 0.55     &      2.42  &       1.06&        1.51   &       3.56  &      1.44  &       2.49  &       0.31  \\
		 &      & $\pm$0.03&  $\pm$0.07 &  $\pm$0.27 &  $\pm$0.27 &  $\pm$0.12 &  $\pm$0.13 &  $\pm$0.15 &  $\pm$0.20 \\
NGC205 f1        & 130  &    ....  &      ....  &       ....&        ....   &       2.08  &      1.44  &       1.78  &       2.87  \\
		 &      &    ....  &      ....  &      ....  &     ....   &  $\pm$0.04 &  $\pm$0.04 &  $\pm$0.05 &  $\pm$0.07 \\
NGC205 f2        & 80   &    ....  &      ....  &       ....&        ....   &       1.57  &      1.46  &       2.25  &       1.60  \\
		 &      &    ....  &      ....  &      ....  &    ....    &  $\pm$0.03 &  $\pm$0.04 &  $\pm$0.04 &  $\pm$0.06 \\
NGC205 f3        & 76   & ....     &      ....  &       ....&        ....   &       1.83  &      1.26  &       2.54  &       3.06  \\
		 &      &    ....  &      ....  &      ....  &     ....   &  $\pm$0.04 &  $\pm$0.04 &  $\pm$0.04 &  $\pm$0.06 \\
NGC205 f4        & 80   & ....     &      ....  &       ....&        ....   &       0.94  &      1.46  &       1.73  &       2.90  \\
		 &      &    ....  &      ....  &      ....  &     ....   &  $\pm$0.04 &  $\pm$0.04 &  $\pm$0.05 &  $\pm$0.06 \\
NGC205 f5        & 54   & ....     &      ....  &       ....&        ....   &       3.02  &      1.46  &       1.39  &       2.27  \\
		 &      &   ....   &      ....  &      ....  &    ....    &  $\pm$0.05 &  $\pm$0.06 &  $\pm$0.07 &  $\pm$0.11 \\
NGC205 f1,f2,f3  & 180  & ....     &      ....  &       ....&        ....   &       1.63  &      1.35  &       2.12  &       2.17  \\
		 &      &  ....    &      ....  &      ....  &    ....    &  $\pm$0.01 &  $\pm$0.01 &  $\pm$0.01 &  $\pm$0.02 \\
NGC205 f4,f5     & 100  & ....     &      ....  &       ....&        ....   &       2.41  &      1.57  &       1.54  &       2.64  \\
		 &      &  ....    &      ....  &      ....  &    ....    &  $\pm$0.03 &  $\pm$0.03 &  $\pm$0.03 &  $\pm$0.04 \\
\hline \hline
\end{tabular}
\end{center}
\end{table*}
\begin{table*}
\begin{center}
\scriptsize
\caption{Globular cluster indices ($ \lambda >$ 4700 \AA) (first line) corrected for zeropoints of
transformation to the standard system (Table~4) and errors
determined from bootstrapping the object spectrum (second line).}
\begin{tabular}{lrrrrrrrr} \\ \hline \hline
GCname         &  H$ \beta$&     Fe5015&       Mg1  &       Mg2   &       Mgb &      Fe5270&     Fe5335&      Fe5406 \\
	       &      (\AA)&      (\AA)&       (mag)&       (mag) &      (\AA)&       (\AA)&      (\AA)&      (\AA)  \\ \hline
HodgeI         &      2.71 &      1.48 &       0.013&       0.076 &       0.67&       1.07 &       0.35&      -0.55   \\
	       & $\pm$0.31& $\pm$0.34& $\pm$0.009& $\pm$0.009& $\pm$0.37& $\pm$0.37& $\pm$0.38& $\pm$0.38\\
HodgeII        &      3.13 &      1.14 &       0.012&       0.029 &       0.62&       0.65 &       0.74&      0.17   \\
	       & $\pm$0.18& $\pm$0.21& $\pm$0.005& $\pm$0.005& $\pm$0.23& $\pm$0.23& $\pm$0.24& $\pm$0.24\\
HodgeIII       &      1.92 &      0.09 &       0.008&       0.031 &       0.32&       0.94 &       0.69&      0.36   \\
	       & $\pm$0.52& $\pm$0.58& $\pm$0.015& $\pm$0.015& $\pm$0.64& $\pm$0.66& $\pm$0.67& $\pm$0.68\\
all N147 GCs   &      2.85 &      1.24 &       0.015&       0.046 &       0.66&       0.97 &       0.61&      -0.02   \\
	       & $\pm$0.11& $\pm$0.12& $\pm$0.003& $\pm$0.003& $\pm$0.14& $\pm$0.14& $\pm$0.14& $\pm$0.14\\
FJJI           &      2.00 &      1.04 &       0.003&       0.077 &       2.09&       0.97 &      -0.13&      0.86   \\
	       & $\pm$0.46& $\pm$0.51& $\pm$0.015& $\pm$0.015& $\pm$0.55& $\pm$0.56& $\pm$0.57& $\pm$0.57\\
FJJIII         &      2.40 &      2.59 &       0.003&       0.057 &       0.79&       0.60 &       0.75&      0.45   \\
	       & $\pm$0.12& $\pm$0.13& $\pm$0.003& $\pm$0.003& $\pm$0.14& $\pm$0.14& $\pm$0.15& $\pm$0.15\\
FJJIV          &      2.70 &      0.61 &      -0.026&       0.018 &       0.78&       0.35 &       0.75&     -0.00   \\
	       & $\pm$0.21& $\pm$0.24& $\pm$0.005& $\pm$0.005& $\pm$0.26& $\pm$0.26& $\pm$0.27& $\pm$0.27\\
FJJV           &      2.73 &      1.94 &       0.010&       0.048 &       0.60&       0.63 &       0.41&      0.37   \\
	       & $\pm$0.10& $\pm$0.11& $\pm$0.002& $\pm$0.002& $\pm$0.12& $\pm$0.12& $\pm$0.12& $\pm$0.12\\
FJJVII         &      2.91 &      2.65 &       0.020&       0.056 &       1.31&       1.85 &       2.16&      0.37   \\
	       & $\pm$0.59& $\pm$0.63& $\pm$0.016& $\pm$0.016& $\pm$0.66& $\pm$0.67& $\pm$0.68& $\pm$0.69\\
FJJVIII        &      2.47 &      2.22 &       0.022&       0.057 &       1.11&       1.15 &      -0.37&      0.32   \\
	       & $\pm$0.51& $\pm$0.54& $\pm$0.014& $\pm$0.014& $\pm$0.59& $\pm$0.60& $\pm$0.60& $\pm$0.60\\
NGC185 field   &      2.08 &      3.65 &       0.053&       0.066 &       0.73&      -0.68 &       1.40&      0.77   \\
	       & $\pm$0.37& $\pm$0.42& $\pm$0.010& $\pm$0.010& $\pm$0.45& $\pm$0.46& $\pm$0.47& $\pm$0.48\\
FJJI+III+IV+   &      2.37 &      1.77 &       0.011&       0.058 &       0.99&       0.90 &       0.02&      0.18   \\
+V+VIII        & $\pm$0.04& $\pm$0.05& $\pm$0.001& $\pm$0.001& $\pm$0.05& $\pm$0.05& $\pm$0.05& $\pm$0.06\\
all N147 GC+   &      2.41 &      1.65 &       0.012&       0.057 &       0.94&       0.85 &       0.13&      0.14   \\
+FJJI+III+IV+  & $\pm$0.04& $\pm$0.04& $\pm$0.001& $\pm$0.001& $\pm$0.04& $\pm$0.04& $\pm$0.04& $\pm$0.04\\
+V+VIII        &           &           &            &             &           &            &           &             \\
HubbleI        &      2.66 &      3.31 &       0.030&       0.067 &       1.29&       1.08 &       1.08&      0.61   \\
	       & $\pm$0.47& $\pm$0.52& $\pm$0.013& $\pm$0.014& $\pm$0.57& $\pm$0.58& $\pm$0.59& $\pm$0.60\\
HubbleII       &      2.32 &      2.75 &       0.017&       0.071 &       1.23&       1.17 &       1.09&      0.68   \\
	       & $\pm$0.05& $\pm$0.06& $\pm$0.002& $\pm$0.002& $\pm$0.06& $\pm$0.06& $\pm$0.06& $\pm$0.06\\
HubbleV        &      4.14 &      3.06 &       0.013&       0.069 &       1.35&       1.69 &       1.31&      0.88   \\
	       & $\pm$0.01& $\pm$0.01& $\pm$0.000& $\pm$0.000& $\pm$0.02& $\pm$0.02& $\pm$0.02& $\pm$0.02\\
HubbleVI       &      3.12 &      2.21 &       0.014&       0.080 &       1.42&       2.51 &       1.88&      0.62   \\
	       & $\pm$0.27& $\pm$0.29& $\pm$0.008& $\pm$0.008& $\pm$0.31& $\pm$0.31& $\pm$0.32& $\pm$0.32\\
HubbleVII      &      2.39 &      2.16 &       0.011&       0.070 &       1.18&       0.94 &       0.45&      0.58   \\
	       & $\pm$0.18& $\pm$0.19& $\pm$0.005& $\pm$0.005& $\pm$0.21& $\pm$0.21& $\pm$0.21& $\pm$0.22\\
HubbleI+VII    &      2.60 &      2.90 &       0.043&       0.081 &       0.62&       0.53 &       0.62&      0.45   \\
	       & $\pm$0.11& $\pm$0.12& $\pm$0.003& $\pm$0.003& $\pm$0.12& $\pm$0.12& $\pm$0.13& $\pm$0.13\\
HubbleVI+FJJVII&      2.86 &      2.82 &       0.016&       0.073 &       1.56&       2.09 &       1.82&      0.72   \\
	       & $\pm$0.21& $\pm$0.22& $\pm$0.006& $\pm$0.006& $\pm$0.24& $\pm$0.24& $\pm$0.25& $\pm$0.25\\
NGC205 f1      &      3.36 &      3.58 &       0.042&       0.091 &       1.74&       0.78 &       1.18&      1.08   \\
	       & $\pm$0.07& $\pm$0.08& $\pm$0.002& $\pm$0.002& $\pm$0.09& $\pm$0.09& $\pm$0.09& $\pm$0.10\\
NGC205 f2      &      3.86 &      4.03 &       0.031&       0.081 &       1.81&       1.67 &       1.34&      1.04   \\
	       & $\pm$0.06& $\pm$0.07& $\pm$0.002& $\pm$0.002& $\pm$0.08& $\pm$0.08& $\pm$0.09& $\pm$0.09\\
NGC205 f3      &      3.60 &      3.25 &       0.018&       0.079 &       1.76&       1.78 &       1.86&      1.42   \\
	       & $\pm$0.06& $\pm$0.07& $\pm$0.002& $\pm$0.002& $\pm$0.08& $\pm$0.08& $\pm$0.09& $\pm$0.09\\
NGC205 f4      &      3.30 &      3.51 &       0.037&       0.081 &       1.53&       1.69 &       1.20&      1.15   \\
	       & $\pm$0.06& $\pm$0.07& $\pm$0.002& $\pm$0.002& $\pm$0.09& $\pm$0.09& $\pm$0.09& $\pm$0.09\\
NGC205 f5      &      2.81 &      2.53 &       0.012&       0.094 &       1.49&       1.05 &       2.02&      0.77   \\
	       & $\pm$0.11& $\pm$0.12& $\pm$0.003& $\pm$0.003& $\pm$0.13& $\pm$0.14& $\pm$0.14& $\pm$0.14\\
NGC205 f1,f2,f3&      3.45 &      3.42 &       0.024&       0.078 &       1.66&       1.62 &       1.37&      1.07   \\
	       & $\pm$0.02& $\pm$0.02& $\pm$0.000& $\pm$0.001& $\pm$0.03& $\pm$0.03& $\pm$0.03& $\pm$0.03\\
NGC205 f4,f5   &      3.12 &      3.21 &       0.029&       0.091 &       1.43&       1.37 &       1.47&      0.93   \\
	       & $\pm$0.04& $\pm$0.05& $\pm$0.001& $\pm$0.001& $\pm$0.06& $\pm$0.06& $\pm$0.06& $\pm$0.06\\
\hline \hline
\end{tabular}
\end{center}
\end{table*}
\begin{figure*}
\vspace{-0.3cm}
\hspace{0.cm}
\includegraphics[width=0.9\textwidth]{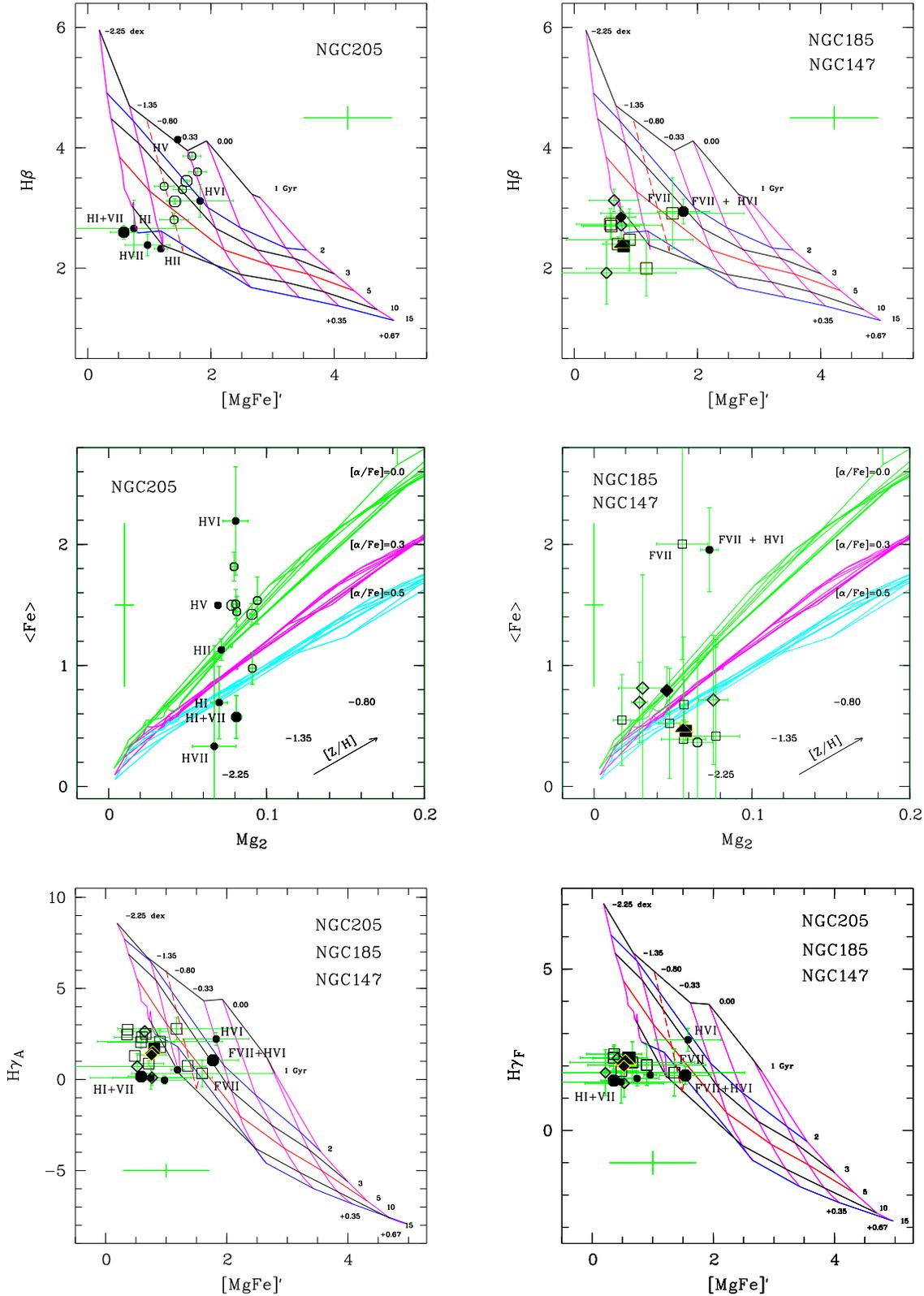}
\vspace{-0.7cm}
\caption{Diagnostic plots for GCs in NGC~205 (solid small circles),
NGC~185 (open squares), NGC~147 (open diamonds), diffuse-light fields (open circles),
and combined spectra of Hubble~I+VII (large filled circle),
FJJVII+HubbleVI (large filled circle),
spectra of FJJI+III+IV+V+VIII (filled square),
all GCs in NGC~147 (filled diamond), all GCs in NGC~147 and NGC~185,
except FJJVII (filled triangle), and combined spectra diffuse-light fields f1+f2+f3
and f4+f5 in NGC~205 (large open circles). The bootstrap errors from Table~5 are overplotted.
We use the SSP model predictions from Thomas et al. (2004).
Error bars on each plot show RMS errors of the transformations to the Lick system (Table~4).}
\label{ind_all}
\end{figure*}

\subsubsection{A three-dimensional interpolation and $\chi^2$ minimization routine}
We developed a program for age, metallicity and \afe\ determination which
involves fitting all available Lick indices using the SSP model
predictions of Thomas et al. (2004). Our three-dimensional linear
interpolation routine obtains a full set of theoretical Lick indices from
required ages, \zh, and \afe\ ratios by minimization of the $\chi^2$
function:
$$ \chi^2 = \sum_{i=1}^{N}\left(\frac{I_i-I_i({\rm
age,\zh,\afe})}{\sigma_{I_i}}\right)^2,$$ where $N$ is the number of Lick
indices involved in the analysis, $I_i$ is an observed index,
$\sigma_{I_i}$ is the total uncertainty of the index, including rms error
of transformations to the Lick/IDS system, $I_i({\rm age,\zh,\afe})$ is
the theoretical index prediction. The program computes 95\% confidence
intervals and 1-$\sigma$ errors of ages, \zh, and \afe. Our routine works
entirely within the three-dimensional space defined by the model grids.
The ranges of estimated values are Age: $[1 \div 15]$ Gyr, $\zh: [-2.25
\div 0.35]$ dex, and $\afe: [0 \div 0.5]$ dex. The program makes
determinations based on 100 combinations of initial conditions randomly
distributed within this three-dimensional space, and the most frequent
event is taken as a value of the global minimum. The errors in age, \zh\
and \afe\ depend primarily on random index uncertainties and uncertainties
of transformations to the Lick/IDS standard system.

To test our routine we compare ages, \zh\ and \afe\ obtained by our
program using the SSP model predictions from Thomas et al. (2004) with the
ones available in the literature for 12 Galactic and 46 M31 globular
clusters (see Figure~\ref{compar}). We use the Lick index data for M31 GCs
with measurement uncertainties $ \Delta H\beta \le 0.5$, $ \Delta
H\gamma_A \le 0.7$, $\Delta H \delta_A \le 0.7$, and $\Delta
[MgFe]\arcmin~ \le 0.4$ \AA\ (see Puzia et al. 2005b) from Worthey et al.
(1994), Kuntschner et al. (2002), Beasley et al. (2004), and Puzia et al.
(2005a). The data on ages, \zh\ and \afe\ for M31 GCs were taken from
Puzia et al. (2005a), and the Lick index data for Galactic globular
clusters from Puzia et al. (2002). When applying our fitting routine to
these Lick index data we use all available indices except for CN1 and CN2.
We found that CN1 and CN2 increase the 1-$\sigma$ errors of our results
and can even skew them (see Proctor et al., 2004). Figure~\ref{compar}
demonstrates linear correlations between our determinations of age,
metallicity and \afe\ and the the corresponding data from the literature.
We found the following linear relations:
$Age_{\rm lit}=Age_{\rm our} + 1.3$ Gyr, $\afe_{\rm lit} = \afe_{\rm
our}-0.02$ dex, $\feh_{\rm lit} = 0.93\cdot\zh_{\rm our}-0.2$ dex for the
sample of MW GCs, and $\zh_{\rm lit} = 1.02\cdot\zh_{\rm our}+0.08$ dex
for M31 GCs.

Only three of the sample Galactic globular clusters have \afe\
measurements from high-resolution spectroscopy of stars in 45 Milky Way
globular clusters by Pritzl et al. (2005). We find that our results for
Galactic GCs are close to the literature data. However, the data for M31
GCs scatter relative to the best least-square fit shown by a solid line on
the third panel in Figure~\ref{compar} with a rms error $\sim$0.2 dex.
More high-resolution spectroscopy of individual stars in young and of
low-\afe\ GCs are needed to determine the real accuracy of our \afe\
determination method. It should be pointed out that the relative error
$\Delta Age/Age$ is fairly constant at around 0.3. Unfortunately, there
are no GCs with young ages and low-\afe\ in our comparison sample to check
for potential biases in our spectroscopic age estimates. The middle panel
of Figure~\ref{compar} shows a tight correlation between \feh\ data from
Harris for Galactic GCs (1996) and our estimates. The existence of the
tight correlation between \feh\ data for Galactic GCs and \zh\ values from
SSP models was also noticed by Thomas et al. (2003) and Proctor et al.
(2004). Proctor et al. (2004) referred this result to problems in the
calibration of the Lick indices or the metallicity scale in SSP models.
According to our exercise the \feh\--\zh\ correlation is not a one-to-one
relation.

\subsubsection{Results}
Resulting ages, metallicities, and \afe\ ratios obtained with our
routine for GCs, diffuse-light fields and the stacked spectra of GCs in
NGC~147, NGC~185, and NGC~205 are listed in Table~7. 1-$\sigma$
uncertainties of our results for each object are summarized in Table~7.

For all studied globular clusters we construct age-metallicity and \afe\
diagnostic plots (see Fig.~\ref{ind_all}) similar to the techniques
described in Puzia et al.~(2005b). It is of interest to see whether the
information coming from the age-metallicity plots coincides well with the
results obtained by applying of our fitting routine. We use the SSP model
predictions of Thomas et al. (2004) and an \afe-insensitive metallicity
index [MgFe]\arcmin~$= \left\{ \mbox{Mg}b \cdot (0.72 \cdot \mbox{Fe5270} +
0.28 \cdot \mbox{Fe5335})\right\}^{1/2}$. 
\begin{table}
\centering
\scriptsize
\caption{Ages, \afe\ and \zh\ for globular clusters and diffuse-light regions for
our sample galaxies. The values were obtained using our fitting routine of
three dimensional interpolation and $\chi^2$ minimization (see Section 2.3
for details). Radial velocities measured by us with a typical error
of about 30 km s$^{-1}$ and approximate projected galactocentric distances
based on the distance estimates from Table~1 are listed in the last two
columns.}
\begin{tabular}{@{}lccccl@{}}
\hline
Object     & Age    & \afe\  & \zh\  & $V_h$ & $D_{proj}$      \\
	   & (Gyr)  &  (dex) &  (dex) & (km/s)&  (kpc)          \\
\hline
{\bf NGC~147}         &              &      &       &                 \\
HodgeI     & 9$\pm$3  & 0.1$\pm$0.3  & $-1.2\pm$0.2  & $-107$& 0.0\\
HodgeII    & 8$\pm$2 & 0.1$\pm$0.3  & $-1.8\pm$0.2  & $-118$& 0.6\\
HodgeIII   & 10$\pm$3 & 0.5$\pm$0.4 & $-1.5\pm$0.2 & $-201$& 0.3\\
all N147 GCs&9$\pm$1 & 0.0$\pm$0.3  & $-1.6\pm$0.1  & & \\
	   &            &             &                &       & \\
{\bf NGC~185} &          &             &         &       & \\
FJJI       & 9$\pm$4  & 0.3$\pm$0.3 & $-1.2\pm$0.2 & $-264$& 0.7\\
FJJIII     & 10$\pm$2  & 0.1$\pm$0.3 & $-1.6\pm$0.3 & $-290$& 0.3\\
FJJIV      & 9$\pm$2   & 0.0$\pm$0.2 & $-2.0\pm$0.3 & $-157$& 0.8\\
FJJV       & 9$\pm$2   & 0.0$\pm$0.3 & $-1.5\pm$0.2 & $-370$& 0.9\\
FJJVII     & 7$\pm$3   & 0.0$\pm$0.3 & $-0.8\pm$0.2 & $-217$& 1.0\\
FJJVIII    & 8$\pm$4   & 0.0$\pm$0.5 & $-1.5\pm$0.3 & $-148$& 1.0 \\
FJJVII+    & 5$\pm$2   & 0.0$\pm$0.3 & $-0.8\pm$0.2 &       &     \\
+HubbleVI  &           &             &                &       &     \\
FJJI+III+&  9$\pm$2& 0.3$\pm$0.2 & $-1.6\pm$0.2   &       & \\
+IV+V+VIII  &           &             &                &       &     \\
field      & 8$\pm$5   & 0.1$\pm$0.4 & $-1.1\pm$0.2   & $-206$& 0.3 \\
all N147 GC+& 10$\pm$2& 0.3$\pm$0.2 & $-1.7\pm$0.1    &       & \\
+FJJI+III+&         &             &                &       & \\
+IV+V+VIII    &           &             &                &       &     \\
{\bf NGC~205}&         &             &                &       & \\
HubbleI    & 7$\pm$2     & 0.2$\pm$0.2 & $-1.1\pm$0.1 & $-144$& 1.0\\
HubbleII   & 10$\pm$2    & 0.0$\pm$0.2 & $-1.2\pm$0.1 & $-235$& 0.7\\
HubbleV    & 1.2$\pm$0.6 & 0.0$\pm$0.1 & $-0.6\pm$0.2 & $-212$& 0.0\\
HubbleVI   & 4$\pm$2     & 0.0$\pm$0.2 & $-0.8\pm$0.2 & $-170$& 0.3\\
HubbleVII  & 11$\pm$2    & 0.1$\pm$0.2 & $-1.3\pm$0.1 & $-189$& 0.4\\
HubbleI+VII& 10$\pm$2    & 0.1$\pm$0.2 & $-1.2\pm$0.1 &       & \\
f1         & 1.9$\pm$0.8& 0.0$\pm$0.2& $-0.5\pm$0.3   & $-245$& 0.05\\
f2         & 1.2$\pm$0.8& 0.0$\pm$0.2& $-0.4\pm$0.3   & $-230$& 0.07\\
f3         & 1.6$\pm$1.1& 0.0$\pm$0.2& $-0.5\pm$0.3   & $-238$& 0.07\\
f4         & 1.8$\pm$1  & 0.5$\pm$0.4& $-0.6\pm$0.3   & $-216$& 0.10\\
f5         & 6$\pm$5    & 0.0$\pm$0.2& $-0.8\pm$0.2   & $-226$& 0.12\\
f1+f2+f3   & 1.6$\pm$0.8& 0.0$\pm$0.2& $-0.5\pm$0.2  &       &     \\
f4+f5      & 4$\pm$2    & 0.0$\pm$0.2& $-0.7\pm$0.2   &       &     \\
\hline
\end{tabular}
\end{table}

Inspection of Fig.~\ref{ind_all} shows that for those globular clusters
with multiple Balmer line indices the ages, metallicities, and \afe\
ratios derived from different diagnostic plots are consistent with each
other. The location of objects in these diagrams correspond well with our
fitting results (Table~7). The problem of age determination for very
metal-poor globular clusters arises, first, since the model lines cross in
this region, and, second, because GCs tend to fall below the oldest model
sequence (see Covino et al. 1995).

Our results reveal solar \afe\ ratios for most of our sample globular
clusters, which is consistent with slow and/or extended globular-cluster/star 
formation histories in their host dSph galaxies. Low \afe\ enhancements were
also found in individual red giant stars in dwarf spheroidal galaxies (dSph) with
high-dispersion spectroscopy (e.g. Shetrone, Bolte \& Stetson 1998,
Smecker-Hane \& McWilliam 1999, Bonifacio et al. 2000, Shetrone et al.
2001, Shetrone et al. 2003, Tolstoy et al. 2003, Geisler et al. 2005).
Shetrone et al. (2001) and Tolstoy et al. (2003) suggested that the low
abundance ratios at low metallicities could result from a slow star
formation rate. Tolstoy et al. suggested that dSph galaxies experienced a
fairly uniform chemical evolution despite their different SF histories.
This conclusion was backed by theoretical model predictions (see e.g.
Lanfranchi \& Matteucci, 2003).

Suppose that our three samples of globular clusters with measured ages,
\zh, and \afe\ ratios in NGC~147, NGC~185, and NGC~205 were drawn from the
same normally-distributed parent populations. To check the hypothesis we
compare our three samples in pairs and apply the F-test (comparison of
variances) and {\it Student's} t-test (comparison of means, see e.g.
Stuart \& Ord 1994). The results of our statistical tests are summarised
in Table~8. The F-test shows that at a significance level of 5\% the
variances are equal. That is the quotient of two variances is always less
than the corresponding critical value of F-distribution. We compute the
T-statistics $ T = t_{1-p} s \sqrt{1/n_1 + 1/n_2}, $ where $t_{1-p}$ is a
critical value of T-distribution at a significance level of $1-p$;
$n_1$ and $n_2$ are numbers of degrees of freedom in the first and the
second samples, and $s$ is a mean weighted variance $s^2= \frac{(n_1 - 1)
s_1^2 + (n_2 - 1) s_2^2}{n_1+n_2-2}$, where $s_1$ and $s_2$ are variances
for the two samples. One can see from Table~8 that the difference between
mean values for the three samples taken in pairs is always less than the
corresponding $T$ value. Hence, at a significance level of 5\% the \afe\
enhancements for the studied globular clusters in NGC~147, NGC~185, and
NGC~205 have similar approximately zero values.

\subsubsection{Individual globular cluster systems}
{\bf NGC 205:} We stack the spectra of the diffuse galaxy in NGC~205 in
order to achieve a higher S/N. The measurements for fields f1--f5 are
indicated by open circles in Fig.~\ref{ind_all}. Large open circles in the
top and middle left panels of Fig.~\ref{ind_all} illustrate the Lick index
values for averaged spectra for the fields of diffuse light in NGC~205:
f1+f2+f3 and f4+f5. Measurements for the combined spectrum for the
clusters HubbleI + HubbleVII are shown as a large filled circle.

It is clear from the H$\beta$ diagnostic plot in the upper left panel of
Fig.~\ref{ind_all} and from Table~7 that the central globular
cluster in NGC~205, Hubble~V, and its surrounding fields f1, f2, f3 are
consistent with the youngest ages in NGC~205 and in our entire sample.
However, the metallicity of Hubble~V appears to be lower than that of the
surrounding fields. On the other hand, the second nearest globular cluster
to the optical center of NGC~205, Hubble~VI, has about the same
metallicity but an older age relative to the very central stellar
population. Finally, the more metal-poor globular clusters Hubble~I,
Hubble~II, and Hubble~VII are the oldest GCs in NGC~205. All three
clusters are located at large galactocentric distances. Hence, our
measurements show that there is a tendency of increasing age with
increasing galactocentric  distance for stellar populations in NGC~205.
This fact is confirmed by extensive CMD studies of stellar populations in
NGC~205 as it will be shown in Section 4.3.

\vspace{0.1cm}
{\bf NGC 185:} This galaxy hosts eight off-center globular clusters. All
of them are located at the projected distance $d_{\rm proj}=$[0.6 -- 1.0]
kpc from the optical center of the galaxy, except FJJIII with $d_{\rm
proj}=0.2$ kpc. Fig.~\ref{ind_all} shows that all observed GCs in NGC~185
are old and metal-poor with the exception of FJJVII, which appears to be
the youngest and most metal-rich globular cluster in NGC~185 (see Tab.~7).
FJJVII has approximately the same age and metallicity as HubbleVI in
NGC~205 according to our measurements. The diagnostic plots and Table~7
show that the stacked spectrum of these two clusters shows the same
intermediate-age and metallicity. Those two globular clusters may have
formed some 4-6 Gyr ago. It will be shown in Sect.~4.2 that
similar intermediate-age and metallicity GCs exist in the halo of M31.

\vspace{0.1cm}
{\bf NGC 147:} All three observed globular clusters in NGC~147 are old and
metal-poor, similar to what is found for globular clusters in NGC~185.
It can be seen in Fig.~\ref{ind_all} that the chemical compositions
obtained from the stacked spectra for all GCs except FJJVII in NGC~185
(filled square), NGC~147 (filled diamond) and GCs Hubble~I, and VII in
NGC~205 (filled circle), as well as for all old and low metallicity GCs in
NGC~185 and NGC~147 (filled triangle), coincide within the errors.
\begin{table}
\begin{center}
\scriptsize
\caption{{\it Students} t-test and  F-test statistics
quantifying the probability
that \afe\ ratios for GCs in NGC~147, 185, and 205 were drawn from the same
distribution. Columns contain the following data: (2) numbers of degrees of freedom,
(3) variances of \afe\ ratios, (4) critical values of F-distribution at a
significance level of 5\% for the three samples taken in pairs, (5) mean \afe\ ratios,
(6) T-statistics, and (7) critical values of T-distribution at a significance level of 5\% for the three samples taken in pairs.}
\begin{tabular}{lrlrccc} \\ \hline
Galaxy    & N & $s^2$ & F$_{0.95}$    & $\langle\afe\rangle$ & T           & t$_{0.95}$ \\ \hline
NGC205 (1)& 5 & 0.008 & 5.0$^{1,2}$ & 0.06                 & 0.11$^{1,2}$& 1.78     \\
NGC185 (2)& 6 & 0.012 & 19.3$^{2,3}$& 0.07                 & 0.19$^{2,3}$& 1.81     \\
NGC147 (3)& 2 & 0.05  & 19.3$^{1,3}$& 0.23                 & 0.17$^{1,3}$& 1.86     \\
\hline
\end{tabular}
\end{center}
\end{table}

\section{Stellar photometry data}
\begin{table*}
 \centering
\caption{Stellar photometry observations. The columns contain the following data:
(1) name of the object (the index "s" marks targets for which spectra were obtained), 
(2) equatorial coordinates in J2000, (3), (4), and (5) source information
from the HST archive, (6) and (7) \feh\ and age obtained by fitting theoretical isochrones 
and ZAHBs of VandenBerg et al. (2006) to our data, (8) the HB morphology index,
(9) search radius inside which we consider stars as members of a particular GCs.}
\begin{tabular}{@{}lcccccccr}
\hline
Object     & RA (J2000) DEC         & Filter, chip & Dataset    & Exposure time      & \feh & age & $\frac{B-R}{B+V+R}$ & $R_{GC}$  \\
	   & \hspace{0.1cm} \fh \hspace{0.2cm} \fm \hspace{0.2cm} \fs \hspace{0.5cm} \fdg \hspace{0.2cm} \farcm \hspace{0.2cm} \farcs  &        &            &  (sec.)              & (dex) & (Gyr) &   & (\arcsec) \\
\hline
{\bf NGC~147}     &                         &                  &        &                &               &    &              &               \\
HodgeII$_s$ & 00 33 15.20 +48 27 23.2 & F555W, F814W, 3  & U2OB01 & 4x1300, 4x1300 & $-2.0\pm$0.1 & 10 & 1:           & 4.8           \\
\noalign{\smallskip}
{\bf NGC~185}     &                         &                  &        &                &            &              &             \\
FJJI$_s$   & 00 38 42.90 +48 18 41.2  & F555W, F814W, PC & U3KL01 & 3x2800, 2x1300 &  $-1.6\pm$0.2 & 10 & 0.2$\pm$0.2& 4.6         \\
FJJII      & 00 38 48.30 +48 18 17.0  & F555W, F814W, PC & U3KL02 & 3x2800, 2x1300 &  $-2.1\pm$0.2 & 10 & $-0.2\pm$0.2& 3.5          \\
FJJIII$_s$ & 00 39 03.90 +48 19 58.2  & F555W, F814W, PC & U3KL03 & 3x2800, 2x1300 &  $-2.0\pm$0.1 & 10 & 0.5$\pm$0.3  & 4.1           \\
FJJIV$_s$  & 00 39 12.38 +48 22 49.2  & F555W, F814W, PC & U3KL04 & 3x2800, 2x1300 &  $-2.0\pm$0.2 & 10 & 0.7$\pm$0.2  & 3.5          \\
FJJV$_s$   & 00 39 13.58 +48 23 05.8  & F555W, F814W, PC & U3KL04 & 3x2800, 2x1300 &  $-1.5\pm$0.1 & 10 & 0.9$\pm$0.1 & 4.0          \\
\noalign{\smallskip}
{\bf NGC~205}     &                         &        &              &                 &              &   &              &            \\
HubbleII$_s$ & 00 40 33.81 +41 39 40.2& F555W, F814W, PC & U3KL06 & 3x2800, 2x1300 &  $-1.3\pm$0.1 & 10& $-0.2\pm$0.3 & 3.2         \\
HubbleVIII & 00 39 53.99 +41 47 19.2  & F555W, F814W, PC & U3KL10 & 3x2800, 2x1300 &  $-1.8\pm$0.2 & 10& 0.7$\pm$0.2  & 3.8         \\
\hline
\end{tabular}
\end{table*}

\subsection{Data Reduction}
\begin{figure*}
\vspace{-3.0cm}
\includegraphics[width=1.1\textwidth]{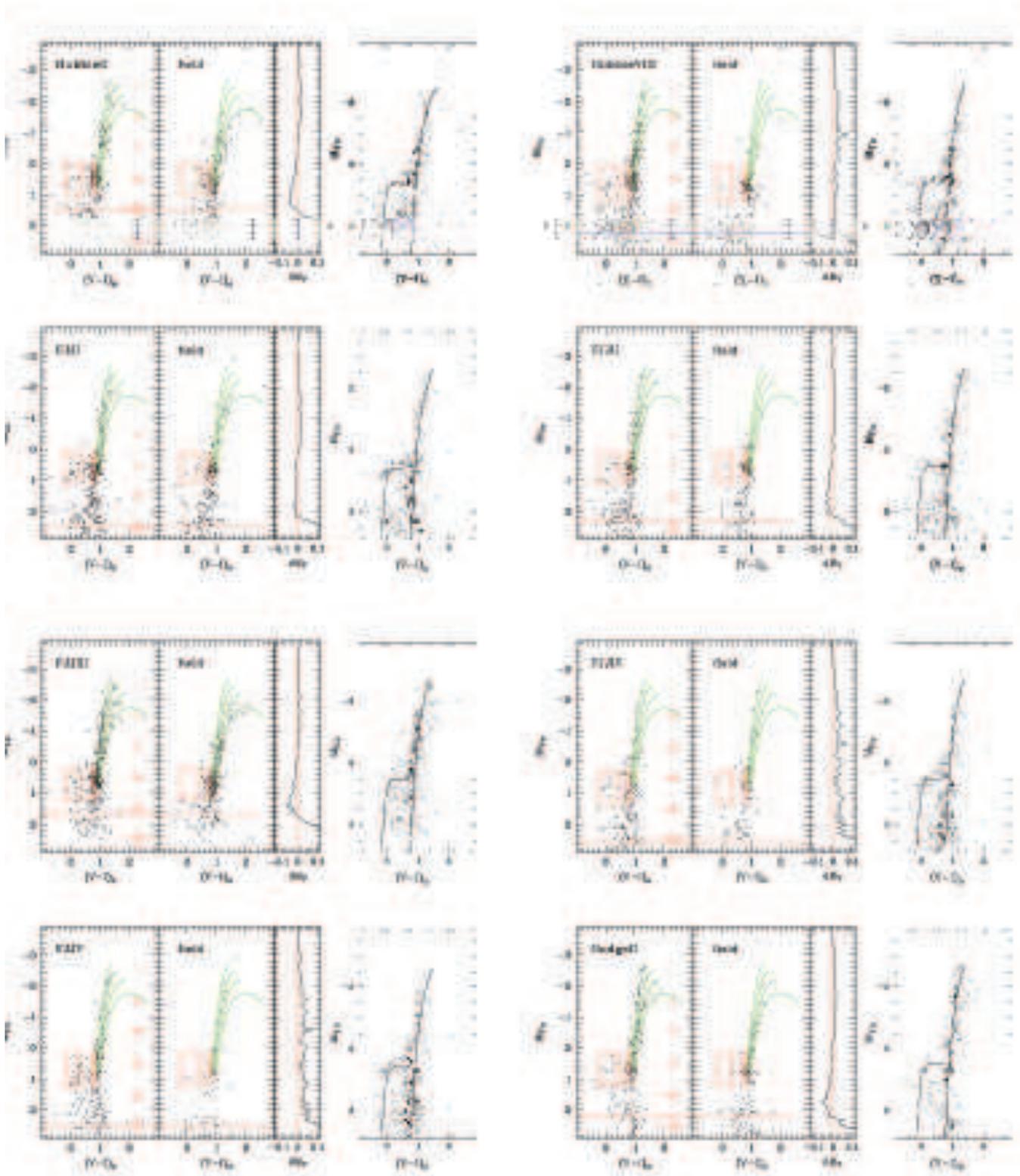}
\vspace{-4.0cm}
\caption{Stellar photometry results.
From left to right 4 panels for each cluster show the following data:
1. uncleaned CMD within the radius of each globular cluster; 2. CMD for
field galactic stars from the equal area; 3. difference between recovered
magnitude and input magnitude for recovered star obtained from HSTphot
artificial star tests; 4. Hess-diagram after decontamination from the
field contribution.
Dashed line on the first and second panels indicates 50\% detection limit.
Empirical loci of the red giant branch (RGB) for Galactic globular
clusters and photometric errors from artificial star tests are shown in
the first panel. The GCs M15, M2, NGC1851, and 47 Tuc have metallicities
\feh~$= -2.17, -1.58, -1.29$, and $-0.71$ dex, respectively (Lee et al.
1993). The BHB and RHB stars in globular clusters are selected according
the boxes. Isochrones and ZAHBs of VandenBerg et al. (2006) for ages,
$\feh$ and $\afe$ indicated in Table~7 are shown in the fourth (right)
panel.}
\label{cmds}
\end{figure*}
In order to test whether our spectroscopic results are consistent with
fundamental parameters derived from CMDs, we search the HST archive for
available photometry for our sample globular clusters. Table~9 summarises
the result of our search, which recovered useful WFPC2 data for eight
globular clusters. We use the HSTphot package (Dolphin, 2000a) to perform
stellar photometry and artificial star tests. We use the charge transfer
efficiency and zero-point magnitude correction derived by Dolphin (2000b).
Stars with S/N~$>5$, $ | \chi | >20$, and $|sharpness|>0.4$, were
eliminated from the final photometry list. The uncertainty of
transformations from the WFPC2 instrumental system to the standard
Johnson-Cousins system contains many factors. The most important of them
are uncertainties of photometric zeropoints ($\sim$0\fm003 for the F555W
filter, $\sim$0\fm002 for F814W filter), the CTE correction ($\sim$0\fm01)
and the aperture corrections ($\sim$0\fm05) (Dolphin, 2000b).

\subsection{Color-Magnitude and Hess Diagrams}
\begin{figure}
\includegraphics[angle=-90,width=0.5\textwidth]{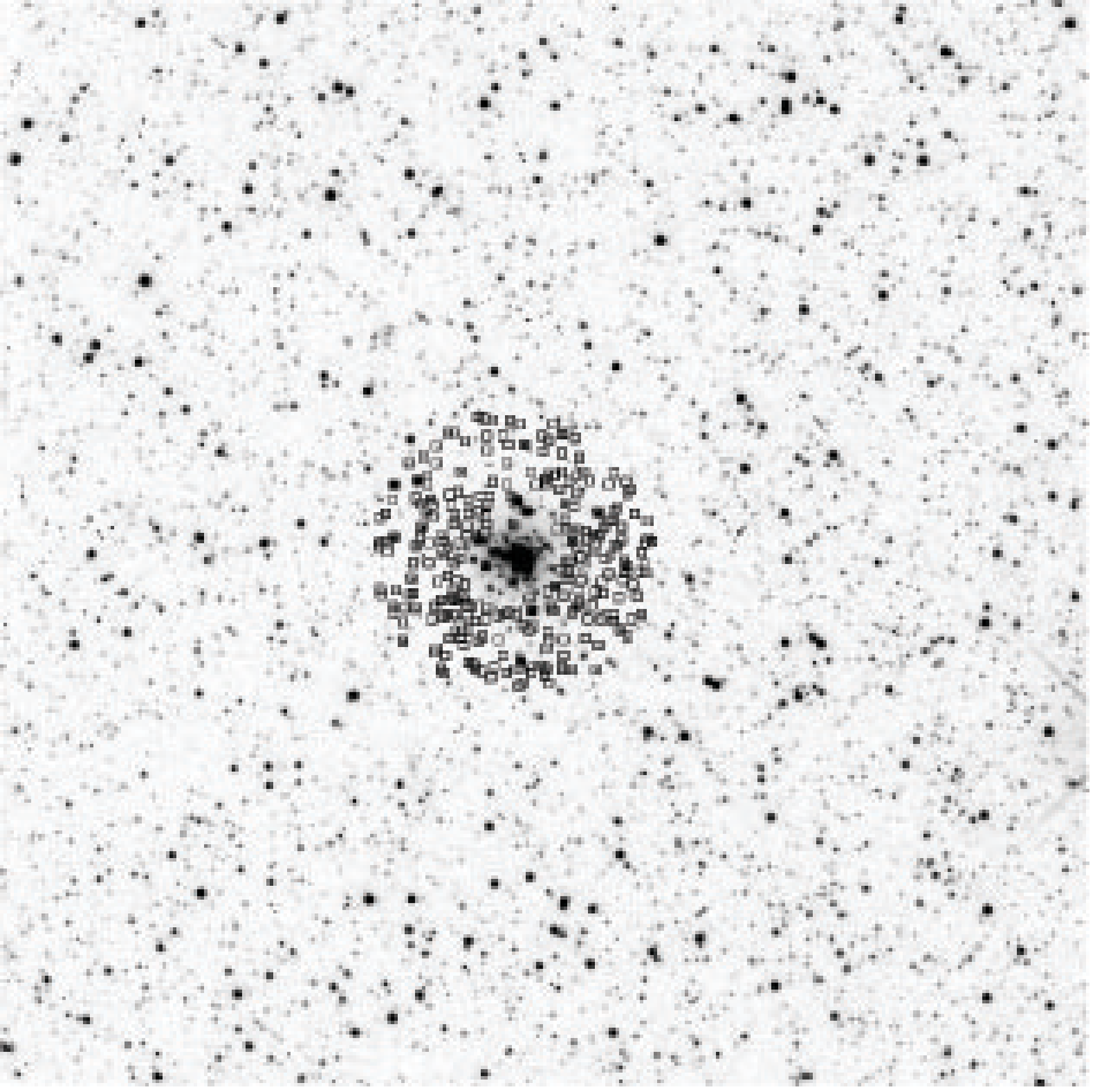}
\includegraphics[angle=-90,width=0.5\textwidth]{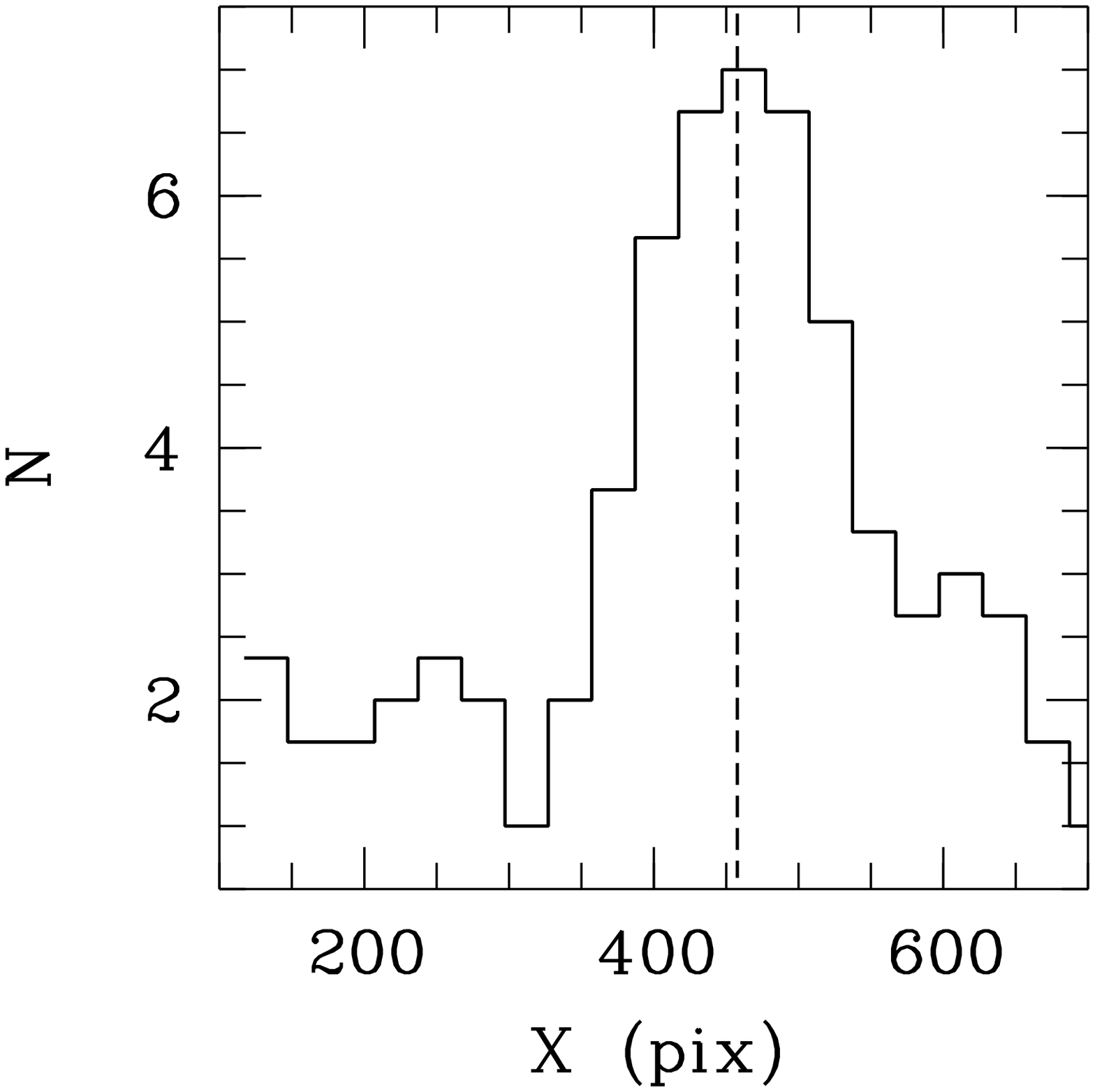}
\caption{ Illustration of our $R_{\rm GC}$ finding technique.
{\it Top panel}: WFPC2/PC image of FJJI ($\sim 34\arcsec\times34\arcsec$,
750x750 pixels). The stars presented in the CMD in Fig.~\ref{cmds} are
marked. {\it Bottom panel}: Distribution of blue stars with $ V-I < 0.2$
in a 200x800 pixel rectangular area centered on the cluster FJJI. The
resulting radius of FJJI is about 100 pixels, which corresponds to
$4\farcs6$.}
\label{surf_distr}
\end{figure}
In order to avoid crowding we analyze in the following only globular
clusters imaged with the PC chip (0\farcs0455/pixel). One exception is the
cluster Hodge~II which is located in a very sparse field of NGC~147 and is
imaged with the WF3 chip of WFPC2. However, the surrounding field of this
cluster is sparse enough to provide reliable stellar photometry. In
general, photometric data for globular clusters imaged with lower
resolution and/or located in the central dense regions of the galaxies
reveal too high photometric errors to confidently determine ages and \feh.
We calculate absolute stellar magnitudes using reddening and distance
estimates listed in Table~1.

Fig.~\ref{cmds} shows CMDs and Hess diagrams for stars within the $R_{GC}$
radius for all our sample GCs. \footnote{The figures for each globular cluster
can be downloaded from the anonymous ftp site: ftp.sao.ru
(directory: /pub/sme/GCsM31dEs).}
We consider the radius of a GC to
be defined by the full area occupied by the bluest stars (old HB stars).
To illustrate how $R_{GC}$ was derived, we demonstrate spatial
distribution of blue stars in the globular cluster FJJI along X axis
(Fig.~\ref{surf_distr}). Blue stars with $V-I<0.2$ were selected in a 200
by 800 pixel rectangular area centered on the cluster. The distribution
shows a clear maximum towards the center of the cluster. Similar
distributions were constructed for each of the studied clusters in X and Y
directions, and the determined two values of $R_{GC}$ were averaged. We
adopt the size roughly 10 -- 20\% smaller than the full area outlined by
blue stars to reduce contamination effects.
 This area roughly corresponds to $R_{GC}$
at a level of the number of blue stars: $N_{back}+ \sigma(N)$, where $N_{back}$ 
is the mean number of blue stars
around the studied cluster, and $\sigma(N)$ is the standard deviation of this number.
Structural parameters of
globular clusters in NGC~205, NGC~185 and NGC~147 will be studied in our
follow-up paper. Here we note that the sizes of the clusters measured
using the distributions of blue stars are $\sim\!3-4$ times larger than
the corresponding effective radii. For example, FJJI has $r_e=1\farcs2$,
whereas its $R_{GC}$ is $\sim\!4\farcs6$. It should be noted that central
regions of the clusters do not show the presence of resolved stars which
pass our goodness-of-fit criteria. This fact was taken into account in
selection of field stars surrounding a cluster. The area for selection of such
stars was computed as $S' = \pi (R_{GC}^2 - R_{c}^2) $, where $R_{GC}$ is
the radius of a globular cluster determined via spatial distribution of
blue stars and $R_{c}$ is the radius of the central unresolved area.

Fig.~\ref{cmds} consists of four panels for each globular cluster. An
uncleaned CMD for stars within $R_{GC}$ of each globular cluster is
illustrated in the left panel. Empirical loci of RGBs for Galactic
globular clusters are shown for comparison (Lee et al., 1993).
CMDs for field stars from an equal area is shown in the second panel. The
third panel represents the difference between recovered magnitude and
input magnitude for recovered stars obtained from artificial star tests. It
is clear that systematic effects do not exceed 0\fm05 for stars with the
completeness limit of $>50$\%, which is indicated by a dotted line. Given the
quality of the data we are able to estimate the horizontal branch
morphology index (see below) with a high level of confidence.

The forth panel represents Hess-diagrams after decontamination from the
field contribution. The statistical field subtraction was done in the
following way: 1. We constructed a Hess-diagram for stars within $R_{GC}$
of each globular cluster. 2. Similar Hess-diagrams were obtained for stars
surrounding the corresponding globular cluster. To take into account
statistical fluctuations of the number of stars we consider stars in an
annulus around the studied cluster. Our simulations show that four
separate areas around the studied cluster are representative enough to
construct Hess-diagrams for field stars, which alleviates the problems of
constructing background samples when a GC is located near to the image
edge. 3. A mean Hess-diagram was computed from these four Hess-diagrams.
4. A cleaned Hess-diagram was calculated as a difference between the
contaminated Hess-diagram and the mean Hess-diagram for field stars.
Positive data indicate the locus of GC stars in the cleaned Hess-diagram.
Solid lines represent the fundamental Zero Age Horizontal Branches (ZAHB)
and isochrones taken from the Victoria-Regina stellar models of VandenBerg
et al. (2006). Corresponding ages and metallicities are listed in Table~9.

\subsection{Comparison with Victoria-Regina Models}
To fit theoretical isochrones to our stellar photometry data we use the
ages, \zh\, and \afe\ obtained in our spectroscopic study.
We note that $\afe = 0$ was adopted for all studied globular clusters
according to our spectroscopic results (see Section 2.3). Pipino et al.
(2006) adopted a relation $\feh = \zh - 0.94 \cdot \afe$ for Thomas et al.
(2003) SSP models on the basis of a calibration on Galactic globular
clusters. Since we found that GCs in NGC~147, 185 and 205 have
approximately solar \afe, it means that $\feh \approx \zh$ in our case. We
have no spectroscopic data for Hubble~VIII and FJJII. As a starting point
for the spectroscopy-CMD comparison we adopt for these two clusters the
age, metallicity, and \afe\ values obtained from the averaged spectra of
Hubble~I+VII and all GCs in NGC~185.
\begin{figure}
\hspace{-0.3cm}
\includegraphics[width=0.36\textwidth,angle=-90]{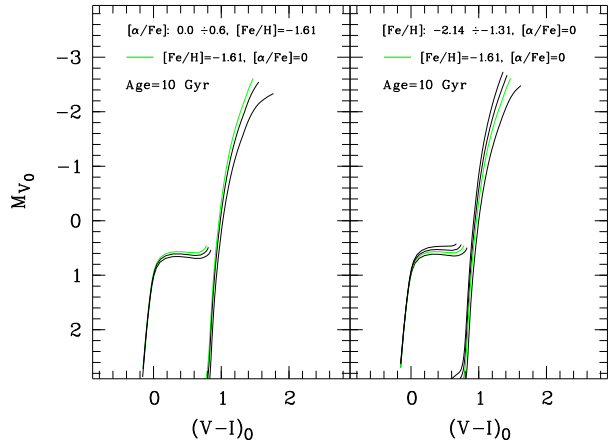}
\caption{Demonstration of the influence of \afe\ and \feh\
on the colour of RGB and the luminosity of ZAHB. We show 
theoretical predictions of the Victoria-Regina stellar models of
VandenBerg et al. (2006) for $\feh=-1.6$ dex, Age=10 Gyr,
and \afe\ ratios 0.0, 0.3, 0.6 (left panel) and
$\feh=$ -2.14, -1.84, -1.61, -1.31, Age=10 Gyr, and $\afe=0$ (right panel).}
\label{VRmodels}
\end{figure}

To show how the choice of different isochrones affects our results we plot
isochrones and ZAHBs (VandenBerg et al., 2006) for different \feh\ and
\afe\ (Fig.~\ref{VRmodels}), and adopt the age of 10 Gyr. The isochrones
and ZAHBs for $Age=10$ Gyr, $\feh =-1.61$ and $\afe =0$ are drawn by grey
lines. The difference in  $(V-I)$ between isochrones that differ by 2 Gyr
is of the order of a few hundredth of a magnitude. So, if we know the
metallicity and \afe, the typical error of age determinations for the
globular clusters studied photometrically is $\sim\!2$ Gyr given the
accuracy of our photometry and the scatter of the data for individual
red-giant stars relative to fitted isochrones. However, since the accuracy
of our spectroscopic metallicity and \afe\ determinations is low in some
cases, the uncertainties of the transformations into the standard VI
system are $\sim$0\fm05, and due to the effect of age-metallicity
degeneracy on the color of the RGB (e.g. Saviane et al. 2000, Dolphin et
al., 2003), we estimate the error of our photometric age determinations to
be $ \la4$ Gyr.

For all GCs, except Hubble~II in NGC~205, the ages, metallicities, and
$\afe$ ratios derived by measuring the Lick indices provide a good
correspondence between the loci of RGB and HB stars and the theoretical
models. The HB stars in Hubble~II seem too bright for their metallicity,
whereas the loci of RGB stars agree well with the spectroscopically
derived metallicity. Our analysis indicates that in order to achieve good
agreement between our spectroscopic and photometric results we have to
reduce the distance modulus to NGC~205, $(m-M)_0=24.58$ (McConnachie,
2005), by 0.2 mag. The resulting distance modulus is given in Table~1. The
same distance modulus was applied to Hubble~VIII in NGC~205.

There is a $\sim\!2\sigma$ discrepancy between our spectroscopic
($\zh=-1.24\pm0.17$) and photometric ($-1.6\pm0.2$) metallicity estimates
for FJJI. We adopt a mean value for this cluster [$\zh=-1.4\pm0.2$ which
coincides with a value obtained by Da Costa \& Mould (1988).

\subsection{Horizontal-Branch Morphology}
Blue HB (BHB) and red HB stars (RHB) in our sample globular clusters were
selected according to the boxes marked in Fig.~\ref{cmds}. The boundaries
of the instability strip in the $V, I$ filter set were taken from Harbeck
et al. (2001): $ (V-I)_0 = 0.4 - 0.75$. We determine the HB morphology
index $ \frac{B-R}{B+V+R} $ (Zinn et al. 1994) for each cluster. The
resulting HB morphology indices are listed in Table~9. Errors of the HB
morphology indices were determined by taking into account variations of a
number of blue and red HB stars in the field regions. All globular
clusters studied here appear to contain a significant population of BHB
stars within their $R_{GC}$ radius.

\section{Discussion}
\subsection{The Second Parameter Effect}
\begin{figure}
\includegraphics[width=0.5\textwidth,angle=-90]{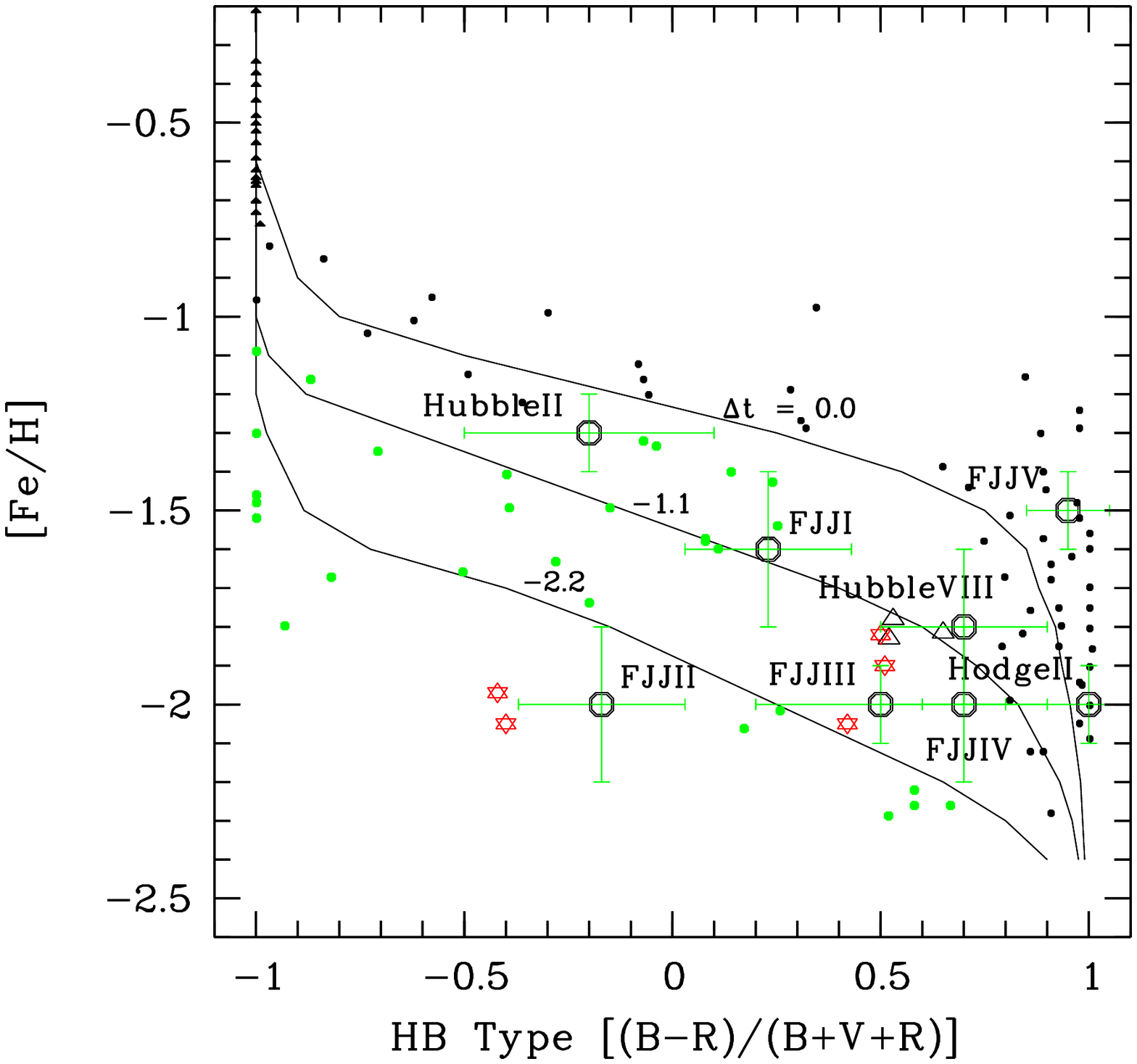}
\caption{HB-type versus metallicity diagram for our sample globular clusters
(open circles), old halo (dots), disk/bulge (triangles) and young halo
Galactic globular clusters (light filled circles), Fornax dSph (stars) and
Sagittarius dSph (open triangles) globular cluster systems. The
overplotted isochrones are from Rey et al. (2001). The ages of the
isochrones are 0.0, $-1.1$, and $-2.2$ Gyr from the top to the bottom.}
\label{hbr}
\end{figure}
The morphology of the HB and the mean luminosity of the HB stars depend
primarily on metallicity. Age is suspected to be the "second parameter" (Lee et
al. 1994). In general, metal-poor old globular clusters have blue HBs and
metal-rich 1--2 Gyr younger clusters have predominantly red HBs. However,
it is known that more parameters may influence the HB morphology such as
helium abundance, CNO abundance etc. (Lee et al.~1994). We suggest that
the photometric data presented in our paper are accurate enough to
consider the influence of age and metallicity on the HB morphology and to
compare this effect with the one studied in the literature for globular
clusters in our Milky Way and other nearby galaxies.

\begin{figure*}
\vspace{-6cm}
\includegraphics[width=0.9\textwidth,angle=0]{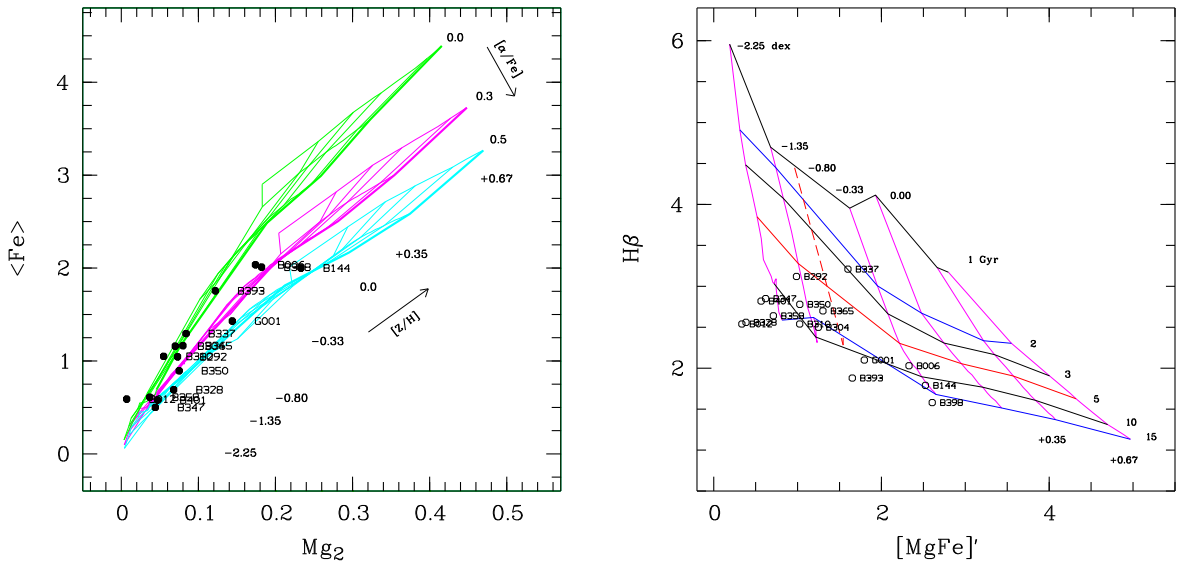}
\vspace{-8cm}
\caption{Age-metallicity diagnostic plots for globular clusters in M31
from a sample of Puzia et al. 2005. We selected clusters with  galactocentric
distances greater than $\sim$40 kpc with the exception of clusters B006 and B012
located near the northern spur (McConnachie et al., 2004) at the
projected distances $\sim$30 kpc and $\sim$28 kpc from the center
of M31, correspondingly.}
\label{ind_M31}
\end{figure*}
Fig.~\ref{hbr} shows the HB-type versus metallicity diagram for our eight
sample globular clusters studied photometrically. The overplotted
isochrones were taken from Rey et al. (2001). It can be seen that the HB
morphologies for six of eight of our sample globular clusters follow the same
behavior as a function of metallicity as the 'young halo' Galactic
globular clusters. Mackey \& Gilmore (2004) performed a comparison between
'old halo', 'young halo', and 'bulge/disc' Galactic globular cluster
sub-populations (Zinn, 1993) using the HB-type versus metallicity diagram.
The authors showed that the majority of external GCs are indistinguishable
from Galactic young halo GCs. Our analysis shows that six of our sample
globular clusters follow a similar trend, typical for 'young halo'
Galactic GCs. We conclude that, as the 'young halo' Galactic GCs, the HB
morphologies for the majority of our sample globular clusters might be, at
least partly, influenced by age, in the sense that our sample GCs have
redder HBs at a given metallicity.

Globular clusters in nearby dwarf galaxies Sagittarius dSph, Fornax dSph,
Magellanic Clouds show similar second-parameter globular clusters which
resemble the 'young halo' Galactic globular clusters in the HB-type versus
$\feh$ diagram (see e.g. Mackey \& Gilmore 2004a,b; Buonanno et al. 1998,
1999; Strader et al. 2003). Rich et al. (2005) studied HB morphologies of
10 GCs in M31 using HST/WFPC2 photometry and found metal-poor globular
clusters with HB morphologies resembling those of the Fornax dSph galaxy
globular clusters. It is of interest to continue the CMD studies using
high resolution HST imaging for GCs in dwarf galaxies and the outskirts of
nearby massive galaxies (e.g. M31) to establish the role of dwarf galaxies
and their globular cluster systems as building blocks of massive galaxies
in the Local group and other nearby groups within 10 Mpc. If most younger
halo GCs in massive galaxies were accreted from low-mass satellites,
globular clusters in dwarf galaxies should resemble these younger halo GCs
in terms of HB type at a given metallicity. However, we do not know
whether these results could be appropriate for probing the formation of
more distant galaxies situated in denser environments like Virgo or Fornax
clusters.

\subsection{Comparison with Globular Clusters in M31}
Perrett et al. 2002, Beasley et al. 2004, Beasley et al. 2005, Fusi Pecci
et al. 2005, and Puzia et al. (2005a) studied GCs in the halo of M31
spectroscopically and found that the M31 globular clusters have
essentially distinct properties from those of the MW globular clusters.
The main conclusions from these studies could be summarised as follows:
1. three types of GC populations exist in the halo of M31: old GCs ($>$10
Gyr) with a wide range of metallicities, intermediate-age GCs with
metallicity $\zh\approx-0.6$, and young GCs ($\leq 1$ Gyr) with slightly
higher metallicities; 2. the M31 GCs have lower \afe\ ratios than the MW
GCs (a mean $\afe=0.14 \pm 0.04$ dex, Puzia et al., 2005a); 3. the
detailed abundances of some elements (e.g. C, N) show a different
behaviour with age and metallicity than that of Galactic GCs.

We select halo globular clusters located at projected galactocentric
distances larger than that of NGC~205 ($\sim40$ kpc) from the samples of
Worthey et al. (1994), Kuntschner et al. (2002), Beasley et al. (2004) and
Puzia et al. (2005a). Fig.~\ref{ind_M31} shows \afe\ and age-metallicity
diagnostic plots for these globular clusters. We expect such
clusters to be the most likely candidates for accreted objects resembling
ages and chemical compositions similar to globular clusters in NGC~147,
NGC~185, and NGC~205. Indeed, ten of sixteen globular clusters selected
from the sample of Puzia et al. (2005a) are old and metal poor and are
located at similar places in the age-metallicity diagnostic plots as our
sample GCs in dwarf galaxies. Only three  of them have $\afe \le 0.0$
according to the data of Puzia et al. (2005a):
B304: ($\zh\sim-1.3$, age $\sim\!13$ Gyr, $\afe\sim-0.12$), B310:
($\zh\sim-1.6$, age $\sim\!12$ Gyr, $\afe\sim-0.33$), B358:
($\zh\sim-2.0$, age $\sim\!10$ Gyr, $\afe\sim0.00$). Some of the selected
M31 globular cluster show high metallicities and old ages (see right panel
of Fig.~\ref{ind_M31}). There are no GCs with such properties in our
sample.

One noteworthy case is the M31 globular cluster B337, which is located at
a projected distance $D_{M31} \sim65$ kpc from the center of M31 near the
northern stellar spur kinematically associated with NGC~205 (McConnachie
et al. 2004). The cluster has an intermediate age of $\sim5$ Gyr and
$\feh\approx-0.6$ dex, and  a solar \afe\ ratio (Puzia et al., 2005a)
similar to that found for Hubble~VI in NGC~205 and FJJVII in NGC~185. 
There are also indication for intermediate-age field stellar populations in the
halo of M31 (e.g. Brown et al. 2003, 2006).

Guhathakurta et al. (2006) discovered an extended halo of metal-poor RGB
stars out to a projected distance of 165 kpc from M31's center. Hence,
NGC~147, NGC~185, and NGC~205 with their globular cluster systems are
embedded in this halo and some clusters may have properties similar to the
M31 halo GCs. Pritzl et al. (2005) demonstrated that most Galactic
globular cluster stars show similar \afe\ ratios as field stars of similar
metallicities, and neither clearly resembles the stellar abundances in
dwarf galaxies. The situation can be more complex for M31, where the
processes of merging and tidal disruption are still active. Additional
spectroscopic and CMD studies of GCs in M31 and dwarf satellites are
needed to establish the fraction of M31 halo GCs which are twins of the
globular clusters in NGC~147, NGC~185, and NGC~205.

%%%%%%%%%%%%

\subsection{The Diffuse Stellar Populations}
Many authors pointed out the large abundance spread across the red giant
branches of our sample galaxies. Mould et al. (1984) found a mean
metallicity for NGC~205 [M/H]$\ga-0.9\pm0.2$ and a metallicity dispersion
$\sigma$[M/H]$\ga0.5$ dex. Additionally, these authors obtained a large
age spread of RGB stars in the range 2--8 Gyr. Mould et al. (1983)
estimated a mean metallicity of RGB stars in the outer parts of NGC~147
[M/H]$=-1.2\pm0.2$ and a metallicity spread of 0.3 dex. Lee et al. (1993)
obtained a metallicity dispersion in NGC~185 in the range $-1.6 \feh<-0.9$
dex from the dispersion of the mean color of RGB stars measured at
$M_I=3.5$. Grebel (2000) derived the following values of metallicity and
the metallicity spread for NGC~205, NGC~185, and NGC~147 correspondingly:
$-0.5 \pm0.5$, $-0.8\pm0.4$ and $-1.1\pm0.4$. Mean metallicities estimated
by McConnachie et al. (2005) for RGB stars in our sample galaxies are
consistent with the data from the aforementioned studies:
[M/H]$_{\alpha=0.0}=-0.8$ for NGC~205, [M/H]$_{\alpha=0.0}=-1.2$ for
NGC185, and [M/H]$_{\alpha=0.0}=-1.1$ for NGC~147.

High resolution HST images made it possible to study RGB stars closer to
the centers of our sample galaxies. Butler and Martinez-Delgado (2005)
identified ancient stars in NGC~185 with $\feh \la-1.5$ dex with the mean
metallicity $-1.11 \pm0.08$ dex and the last period of active star
formation dating about 4x$10^8$ yr ago. The most metal-rich RGB stars in
NGC~205 were found to reach $\feh \ga -0.7$ dex, while a median value of
metallicity for ancient stars in this galaxy is $\feh=-1.06 \pm0.04$.
Dolphin (2005) using the same observational material and the DOLPHOT
program obtained mean metallicities $-0.6, -0.9,$ and $-0.9$ dex for
NGC~205, NGC~185, and NGC~147, respectively.

Our spectroscopic study deals with the integrated light from stars in
NGC~205 and NGC~185 very near to the centres of the host galaxies. Hence,
it is worth to note that our study supplements the results obtained by the
extensive CMD studies. We found $\overline{\zh} \sim 0.5$ dex for the
central regions of NGC~205 and a large spread of metallicity $\sim 0.3
\div 0.4$ dex which could not be explained by measurement uncertainties,
because the spread is considerably less for globular clusters having the
same S/N ratio in their spectra. We estimate the metallicity of the region
of diffuse galactic light in NGC185 near the globular cluster FJJIII
($d_{proj}=0.2$ kpc) to be $\zh=-1.16 \pm 0.25$ in a good consent with
the CMD studies. As it was also mentioned by Da Costa \& Mould (1988) the
mean abundance of the associated globular clusters in each of the studied
galaxies is considerably less than that observed for the corresponding
field halo stars. So, one could suggest that globular cluster systems of
these galaxies formed at the earliest epochs of galactic formation along
with the first stars.

We found indications for increasing age with increasing galactocentric
distance for the field stellar population in NGC~205. The same trend
was found in deep CMD studies for stellar populations in NGC~205 and
NGC~185. According to Cappellari et al. (1999) there is a central
population of stars in NGC~205 with ages in the range 50 -- 100 Myr. AGB
stars have solar metallicity and are distributed uniformly over the inner
$\sim\!1\arcmin$ (Davidge~1992). Old stars in NGC~185 form the less
concentrated system (Martinez-Delgado et al.~1999). The youngest stars are
located in the central 150x90 parsec$^2$ region (Butler \&
Martinez-Delgado, 2005). Outside this area only stars with ages $\ge 1$
Gyr are found.

%%%%%%%%%%%%%%%%%%%%%%%%%%%%%%%%
\section{Conclusion}
 Although the modest S/N of some of our GC spectra and the relatively
moderate photometric depth of our CMDs do not allow us to derive very
accurate ages, we conclude that all our sample GCs appear to be old ($T>8$
Gyr) and metal-poor ($\zh \la -1.1$), except for the GCs Hubble~V in
NGC~205 ($T=1.2\pm0.6$ Gyr, $\zh=-0.6\pm0.2$), Hubble~VI in NGC~205
($T=4\pm2$ Gyr, $\zh =-0.8\pm0.2$), and FJJVII in NGC~185 ($T=7\pm3$ Gyr,
$\zh =-0.8\pm0.2$). We find two intermediate-age GCs (HubbleVI and FJJVII)
which are located at projected distances $D_{proj} \sim0.3$ and 1.0 kpc
from the centers of their host galaxies. Deep images taken with HST/ACS
are necessary to understand whether these relatively faint star clusters
are genuine globular clusters (i.e. simple stellar populations) born during
$\sim\!4-6$ Gyr old star-formation events in their host galaxies.
They could consist also of multiple stellar populations similar to
$\omega$Cen (see e.g. Sollima et al. 2005).

The HB morphologies for our sample GCs follow the same behavior with
metallicity as younger halo Galactic globular clusters. It was found for
Galactic GC, that age may not be the dominant second parameter determining
the shape of globular clusters' HBs (see e.g. Stetson et al., 1996). GCs
having blue HBs are in general older than 10 Gyr (Lee et al. 1994). So, we
suggest here that the HB morphologies for our sample GCs likely do not
bias our spectroscopic age estimates based on Balmer absorption line
indices.

We find that most of the GCs in the studied galaxies are weakly or not
$\alpha$-enhanced, in contrast to the population of GCs in nearby
early-type galaxies (see Puzia et al. 2006). The chemical composition of
globular clusters may turn out to be a powerful tool to discriminate
between clusters which formed {\it in situ} in massive galaxies and those
that were formed in smaller sub-units and later accreted in more massive
halos.

Spectroscopic ages and metallicities of the central regions in NGC~205 and
NGC~185 coincide with those obtained from color-magnitude diagrams. The
central field stellar populations in these galaxies have approximately the
same age as their most central GCs Hubble~V in NGC~205 and FJJIII in
NGC~185, respectively, but are more metal-rich than the central globular
clusters.

\section*{Acknowledgments}
MES acknowledges all participants of the project SCORPIO at scientific
advisory of V.L.~Afanasiev for their extensive work on designing,
producing and testing the SCORPIO spectrograph with the multislit unit at
the 6m telescope. THP is supported by an ESA Research Fellowship, which is
gratefully acknowledged. THP also acknowledges partial financial support
through grant GO-10129 from the Space Telescope Science Institute, which
is operated by AURA, Inc.,~under NASA Contract NAS5-26555. MES thanks Don
VandenBerg for his help in applying his models to the analysis of the CMDs
for GCs in NGC~205, N.A.~Tiknonov, D.I.~Makarov, O.K.~Sil'chenko,
O.A.~Galazutdinova, G.M.~Beskin. We thank the anonymous referee for
valuable comments. VLA acknowledges the INTAS grant (96-0315), the
"Astronomy" Federal Science and Technology Program (contract no.
40.022.1.1.1101 from February 1, 2002) and the Program of the Department
of Physical Sciences of the Russian Academy of Sciences for partial
support of his work. The 6-m telescope of the Special Astrophysical
Observatory of the Russian Academy of Sciences is operated under the
financial support of the Science Department of Russia (registration number
01-43). Some of the data presented in this paper were obtained from the
Multimission Archive at the Space Telescope Science Institute (MAST).
STScI is operated by the Association of Universities for Research in
Astronomy, Inc., under NASA contract NAS5-26555. Support for MAST for
non-HST data is provided by the NASA Office of Space Science via grant
NAG5-7584 and by other grants and contracts. This research has made use of
NASA's Astrophysics Data System Bibliographic Services.

\end{document}